\documentclass[onecolumn,showpacs,amsmath,amssymb]{revtex4}
\input{epsf}
\usepackage{graphicx}
\usepackage{array}
\usepackage{dcolumn,longtable}

\usepackage{rotating,booktabs}
\usepackage{booktabs,threeparttable}
\begin{document}

\title{Non-relativistic ab initio calculations 
for $2\,^2\!S$, $2\,^2\!P$ and $3\,^2\!D$ lithium isotopes: Applications to polarizabilities and  dispersion interactions}
\author{Li-Yan Tang$^{1,2}$, Zong-Chao Yan$^{3,4}$, Ting-Yun Shi$^{1}$, and James F. Babb$^{5}$}

\affiliation {$^1$State Key Laboratory of Magnetic Resonance and
Atomic and Molecular Physics, Wuhan Institute of Physics and
Mathematics, Chinese Academy of Sciences, Wuhan 430071, China}

\affiliation {$^{2}$Graduate School of the Chinese Academy of
Sciences, Beijing 100049, China }

\affiliation{$^3$Department of Physics, University of New Brunswick,
Fredericton, New Brunswick, Canada, E3B 5A3}

\affiliation{$^4$Center for Theoretical Atomic and Molecular
Physics, the Academy of Fundamental and Interdisciplinary Sciences,
Harbin Institute of Technology, Harbin 150080, China}

\affiliation{$^5$ITAMP, Harvard-Smithsonian Center for Astrophysics, Cambridge, Massachusetts 02138, USA}

\date{\today}

\begin{abstract}
{The  electric dipole polarizabilities and hyperpolarizabilities for the lithium
isotopes ${}^6\textrm{Li}$ and  ${}^7\textrm{Li}$ in
the ground state $2\,^2\!S$ and the excited states $2\,^2\!P$ and $3\,^2\!D$,
as well as the leading resonance and dispersion long-range coefficients for the
Li($2\,^2\!S$)--Li($2\,^2\!S$) and Li($2\,^2\!S$)--Li($2\,^2\!P$) systems,
are calculated nonrelativistically using variational wave
functions in Hylleraas basis sets.
Comparisons are made with published results, where available.
We  find that the value of the second hyperpolarizability of the $2\,^2\!S$ state
is sensitive to the isotopic mass due
to a near cancellation between two terms.
For the $3\,^2\!D$ state polarizability tensor
the calculated components disagree  with  those measured in
the sole experiment and with those calculated  semi-empirically.
}
\end{abstract}

\maketitle

\section{Introduction}

The energies, transition probabilities, and polarizabilities of the lithium atom and its isotopes
and their mutual long-range interaction coefficients
have been extensively studied, and many calculational approaches have been
developed and tested against each other and against available experimental results.
Nevertheless---as we shall show---there is still a need for calculations of increasingly
high-precision to serve as benchmarks, to predict atomic and interatomic coefficients,
and to help understand
discrepancies  between various results.
Calculations at the level reflecting the
mass of lithium isotopes may be applied to a  diverse set of
recent areas of interest in, for example, astrophysics~\cite{AspLamNis06},
ultracold atom-atom scattering~\cite{AbrMcAGer97,BarAltRie05}
and Feshbach resonance analyses~\cite{vanMarKok04,PolDriJun09},
photoassociation spectroscopy~\cite{McAAbrHul96,SchDeuSil03},
atom-molecule scattering~\cite{CviSolHut07}
and three-atom inelastic collisional loss studies~\cite{NaiUed08}.
Measurements
and calculations on excited states and their properties are also at the frontier, though
discrepancies between  theory and experiment remain
even for the lowest fine structure levels~\cite{YanNorDra08,PucPac09}.

It has been demonstrated that \textit{ab initio} wave functions
obtained  variationally using Hylleraas-type basis functions are
capable of yielding highly accurate results for Li properties,
\textit{cf.}~\cite{yan,Kin97}. In turn, those results have proven
useful in gauging the effectiveness of Gaussian-type basis
functions~\cite{StaKomKed08} and other
calculations~\cite{GodFroFis01}, semi-empirical
methods~\cite{ZhaMitBro07}, and relativistic
approaches~\cite{JohSafDer08,WanSahTim08,DerPorBel08}. The purpose
of the present work is to apply and extend methods developed over a
series of previous
papers~(\textit{e.g.}~\cite{yan,YanDra95b,yan-drake-97}) to the
excited  $2\,^2\!P$ and $3\,^2\!D$ states of the lithium isotopes
thereby providing a consistent  and highly accurate \textit{ab
initio} treatment of the polarizabilities and their related
quantities using the nonrelativistic Schr\"{o}dinger equation. We
also refine the previous results~\cite{yan} for the $2\,^2\!S$
ground state by improving the accuracy and by the inclusion of the
isotope mass. For the excited states of the isotopes we provide
calculations of static polarizabilities and static second
hyperpolarizabilities and we investigate the  excited state
resonance and dispersion long-range interaction potential energy
coefficients.

\section{Theory}\label {theory}

In this section, the Hamiltonian and basis sets
will be exhibited along with
expressions for the electric multipole transition operators, polarizabilities,
second hyperpolarizabilities,
and dispersion coefficients.
The formulation extends and unifies those given previously
for Li~\cite{yan,YanDra95b,yan-drake-97}, for He~\cite{ZhaYanVri06a}, and for
$\textrm{H}_2^+$~\cite{zhang-yan04};
we include the isotopic mass and we treat the response
of the atom to an applied electric field.

\subsection{Hamiltonian}\label {hamiltonian}

The  transformation from the laboratory frame to the center of
mass frame that we will use for describing the Li atom was
given by Zhang and Yan~\cite{zhang-yan04} in  a general form for $n+1$
charged particles.  It was applied to generate the Hamiltonian
and transition operators for $\textrm{H}_2^+$ in Ref.~\cite{zhang-yan04}
and for He in Ref.~\cite{ZhaYanVri06a}.

We directly follow the expressions given in
Eqs.~(20)--(26) of Ref.~\cite{zhang-yan04},
where the particles are labeled by the index  $i=0,...,n$.
We make  the replacement $n=3$, identify
particle $0$ with the nucleus of mass $m_0$,
identify the particles $i=1,2$ and $3$ with the electrons,
so that $m_1=m_2=m_3=m$, where $m$ is the electron mass,
to obtain the Hamiltonian expressed in
the center of mass frame describing the internal motion of
the Li atom,
\begin{eqnarray}
H_0 &=& -\frac{1}{2\mu}\sum_{i=1}^3 \nabla_i^2
-\frac{1}{m_0}\sum_{i> j\ge 1}^3\nabla_i\cdot\nabla_j
+q_0\sum_{i=1}^3\frac{q_i}{r_i} +\sum_{i> j\ge
1}^3\frac{q_iq_j}{r_{ij}}\,, \label {eq:t6}
\end{eqnarray}
where $\mathbf{r}_i$ is the position
vector of electron $i$ from the
nucleus, and  $\mu=m m_0/(m +m_0)$ is the reduced mass.
In this subsection $j$ stands for a summation index (in
subsequent parts of the paper it will
be an angular momentum quantum number), and
$q_j$, $j=0,...,3$ are the charges of the respective particles.

The  $2^\ell$-pole transition
operator is
\begin{eqnarray}
T_\ell &=& \sum_{i=0}^3 q_i R_i^\ell Y_{\ell 0}({\hat{\bf R}}_i)\,,
\label {eq:t7}
\end{eqnarray}
where $\mathbf{R}_i$, $i=0,...,3$,
as shown in
Eq.~(27) of Ref.~\cite{zhang-yan04}.
It can be transformed
into the center of mass frame by applying Eqs.~(23) and (24)
of Ref.~\cite{zhang-yan04}, which may be written in the form
\begin{eqnarray}
{\bf R}_i  &=&  \sum_{j=1}^3 \epsilon_{ij} {\bf r}_j\,, \label
{eq:t8}
\end{eqnarray}
with $\epsilon_{ij}=\delta_{ij}-m_j/M_T$, $i=0,1,2,3$, $j=1,2,3$,
and $M_T=m_0+3m$. The general formula for $T_{\ell}$ expressed in
the center of mass frame is given in Eq.~(30) of
Ref.~\cite{zhang-yan04}.

For a four-particle system,
it can be shown that the transition operators $T_\ell$ with $\ell$
up to 3 have the following explicit forms:
\begin{eqnarray}
T_1 &=& \sum_{j=1}^3\bigg(\sum_{i=0}^3 q_i\epsilon_{ij}\bigg) r_j
Y_{10}({\hat{\bf r}}_j)\,, \label {eq:t11}
\end{eqnarray}
\begin{eqnarray}
T_2 &=& \sum_{j=1}^3\bigg(\sum_{i=0}^3 q_i\epsilon_{ij}^2\bigg)r_j^2
Y_{20}(\hat{\bf r}_j) + \sqrt{\frac{15}{2\pi}} \sum_{\{j,k\}}
\bigg(\sum_{i=0}^3q_i\epsilon_{ij} \epsilon_{ik}\bigg)r_j
r_k(\hat{\bf r}_j\otimes \hat{\bf r}_k)^{(2)}_0 \,, \label {eq:t12}
\end{eqnarray}
with $\{j,k\}=(1,2)$, $(2,3)$, and $(3,1)$, and
\begin{eqnarray}
T_3 &=& \sum_{j=1}^3\bigg( \sum_{i=0}^3
q_i\epsilon_{ij}^3\bigg)r_j^3 Y_{30}(\hat{\bf r}_j)
      +3\sqrt{\frac{35}{8\pi}}
\sum^3_{\substack{
j,k=1 \\
j\ne k }} \bigg(\sum_{i=0}^3 q_i\epsilon_{ij}^2 \epsilon_{ik} \bigg)
r_j^2r_k((\hat{\bf r}_j\otimes\hat{\bf r}_j)^{(2)} \otimes\hat{\bf
r}_k)^{(3)}_0\nonumber \\
&+&6\sqrt{\frac{35}{8\pi}}\bigg(\sum_{i=0}^3
q_i\epsilon_{i1}\epsilon_{i2}\epsilon_{i3} \bigg)r_1r_2r_3((\hat{\bf
r}_1\otimes\hat{\bf r}_2)^{(2)}\otimes\hat{\bf r}_3)^{(3)}_0 \,.
\label {eq:t13}
\end{eqnarray}

For a neutral system, the finite mass effect enters into the transition
operator $T_\ell$ as a polynomial of degree $\ell-1$ in $m_j/M_T$.
For Li, the three coefficients appearing
in the sets of  parentheses in Eq.~(\ref{eq:t13}) for the
transition operator $T_3$, for example, can be written as
\begin{eqnarray}
\sum_{i=0}^3 q_i\epsilon_{ij}^3 &=&
q_j-3q_j\bigg(\frac{m_j}{M_T}\bigg)+3q_j\bigg(\frac{m_j}{M_T}\bigg)^2\,, \label{eq:t14}\\
\sum_{i=0}^3 q_i\epsilon^2_{ij}\epsilon_{ik} &=&
-q_j\bigg(\frac{m_k}{M_T}\bigg)+2q_j\bigg(\frac{m_j}{M_T}\bigg)\bigg(\frac{m_k}{M_T}\bigg)
+q_k\bigg(\frac{m_j}{M_T}\bigg)^2\,,
\label{eq:t15}\\
\sum_{i=0}^3 q_i\epsilon_{i1}\epsilon_{i2}\epsilon_{i3} &=&
q_1\bigg(\frac{m_2}{M_T}\bigg)\bigg(\frac{m_3}{M_T}\bigg)
+q_2\bigg(\frac{m_3}{M_T}\bigg)\bigg(\frac{m_1}{M_T}\bigg) +
q_3\bigg(\frac{m_1}{M_T}\bigg)\bigg(\frac{m_2}{M_T}\bigg)
\,. \label{eq:t16}
\end{eqnarray}
For an atomic system of
infinite nuclear mass, all the coefficients are zero except for those
of $r_j^\ell Y_{\ell 0}(\hat{\bf r}_j)$,  which are equal to
$q_j$, as expected.
We now let $m=1$ to utilize atomic units for the remainder of the paper.

\subsection{Variational basis sets}\label {basis}

The wave functions are
obtained by solving variationally the energy eigenvalue equation for lithium
\begin{eqnarray}
H_0\Psi_0( {\bf{r}}_{1},{\bf{r}}_{2},{\bf{r}}_{3}) &=& E_0\Psi_0(
{\bf{r}}_{1},{\bf{r}}_{2},{\bf{r}}_{3})\,, \label {eq:t17}
\end{eqnarray}
with $H_0$ given by Eq.~(\ref{eq:t6}),
in terms of the explicitly correlated basis functions in Hylleraas
coordinates,
\begin{eqnarray}
\phi({\bf{r}}_{1},{\bf{r}}_{2},{\bf{r}}_{3})=r_{1}^{j_{1}}r_{2}^{j_{2}}r_{3}^{j_{3}}r_{12}^{j_{12}}r_{23}^{j_{23}}r_{31}^{j_{31}}e^{-\alpha
r_{1}-\beta r_{2}-\gamma r_{3}}
\mathcal{Y}_{(\ell_{1}\ell_{2})\ell_{12},\ell_{3}}^{LM_{L}}(\hat{{\bf{r}}}_{1},\hat{{\bf{r}}}_{2},
\hat{{\bf{r}}}_{3})\chi(1,2,3)\,, \label {eq:t18}
\end{eqnarray}
where $r_{ij}=|{\bf{r}}_{i}-{\bf{r}}_{j}|$ is the inter-electronic
separation,
$\mathcal{Y}_{(\ell_{1}\ell_{2})\ell_{12},\ell_{3}}^{LM_{L}}$ is a
vector-coupled product of spherical harmonics to form an eigenstate
of total angular momentum $L$ and component $M_L$, which can be
written in the form
\begin{eqnarray}
&&\mathcal{Y}_{(\ell_{1}\ell_{2})\ell_{12},\ell_{3}}^{LM_{L}}(\hat{{\bf{r}}}_{1},
\hat{{\bf{r}}}_{2},\hat{{\bf{r}}}_{3})
=\sum_{{\rm all\,} m_{i}}\langle
\ell_{1}m_{1};\ell_{2}m_{2}|\ell_{1}\ell_{2};\ell_{12}m_{12}\rangle
\nonumber\\
&&\times
\langle\ell_{12}m_{12};\ell_{3}m_{3}|\ell_{12}\ell_{3};LM_{L}\rangle
Y_{\ell_{1}m_{1}}(\hat{{\bf{r}}}_{1})
Y_{\ell_{2}m_{2}}(\hat{{\bf{r}}}_{2})Y_{\ell_{3}m_{3}}(\hat{{\bf{r}}}_{3})\,,
\label {eq:t19}
\end{eqnarray}
and $\chi(1,2,3)$ is the three-electron spin $1/2$ function. The
variational wave function $\Psi_0$ is a linear combination of basis
functions $\phi$, anti-symmetrized.
The procedures followed are similar to those described in Ref.~\cite{YanDra95b}.
With some truncations to avoid
potential numerical linear dependence, all terms in
Eq.~(\ref{eq:t18}) are included such that
\begin{eqnarray}
j_1+j_2+j_3+j_{12}+j_{23}+j_{31} \le \Omega\,, \label {eq:t20}
\end{eqnarray}
where $\Omega$ is an integer, and the convergence for the energy
eigenvalue is studied by increasing $\Omega$ progressively. The
basic type of integral that appears in this work is of the form
\begin{eqnarray}
&&\int d{\rm\bf r}_1 d{\rm\bf r}_2 d{\rm\bf r}_3 \,
r_1^{j_1}r_2^{j_2}r_3^{j_3}
r_{12}^{j_{12}}r_{23}^{j_{23}}r_{31}^{j_{31}}
e^{-\alpha r_1-\beta r_2-\gamma r_3}\nonumber\\
&&\times Y_{\ell'_1m'_1}^{*}(\hat{{\bf r}}_1)
Y_{\ell'_2m'_2}^{*}(\hat{{\bf r}}_2) Y_{\ell'_3m'_3}^{*}(\hat{{\bf
r}}_3) Y_{\ell_1m_1}(\hat{{\bf r}}_1) Y_{\ell_2m_2}(\hat{{\bf
r}}_2)Y_{\ell_3m_3}(\hat{{\bf r}}_3)\,  \label {eq:t21}
\end{eqnarray}
and computational details for this integral can be found in
Ref.~\cite{yan-drake-97}.

\subsection{Stark effect and polarizabilities}\label {stark}

The polarizability of an atom can be considered as a
measure of the response of the charge cloud to an external electric field,
which can be illustrated by the Stark effect. Consider a lithium
atom in a weak external electric field $\boldsymbol{\cal{E}}=\cal
{E}\, \bf{\hat{z}}$. The initial state is assumed to be a parity
eigenstate and is written as $|0\rangle\equiv|n_0 LM\rangle $,
where $n_0$ is the principal quantum number and $L$ and $M$ are the
usual angular momentum quantum numbers. According to the
perturbation theory, the energy shift due to $\boldsymbol{\cal{E}}$
can be expressed in the form
\begin{eqnarray}
\Delta E &=& \Delta E_2+\Delta E_4\,,
 \label {eq:t22}
\end{eqnarray}
where $\Delta E_2$ and $\Delta E_4$ are, respectively,  from the second- and
fourth-order corrections, whereas the first- and
third-order corrections are zero because of the parity selection
rule. The detailed derivation for $\Delta E_2$ and $\Delta E_4$ is
given in the Appendix. The final expression for $\Delta E_2$ is
\begin{eqnarray}
\Delta E_2 &=&
-\frac{{\cal{E}}^2}{2}\bigg[\alpha_1+\alpha_1^{(T)}g_2(L,M)\bigg]\,,
 \label {eq:t23}
\end{eqnarray}
where $g_2(L,M)$ is the only $M$-dependent part, defined in the Appendix by Eq.~(\ref{eq:a20}),
and $\alpha_1$ and $\alpha_1^{(T)}$ are,  respectively,  the scalar and tensor dipole
polarizabilities. The polarizabilities $\alpha_1$ and $\alpha_1^{(T)}$ can
be expressed further  in terms of the reduced matrix elements of the dipole
transition operator:
\begin{eqnarray}
\alpha_1 &=& \sum_{L_a} \alpha_1(L_a)\,, \label{eq:t24}\\
 \alpha_1^{(T)} &=& \sum_{L_a}
W(L,L_a)\alpha_1(L_a)\,, \label {eq:t25}
\end{eqnarray}
where
\begin{eqnarray}
\alpha_1(L_a) &=& \frac{8\pi}{9(2L+1)}\sum_n \frac{|\langle n_0 L
\|T_1\| nL_a\rangle |^2}{E_n(L_a)-E_{n_0}(L)}\,,
 \label {eq:t26}
\end{eqnarray}
with $T_1=\sum_{i=0}^3 q_iR_i Y_{10}({\bf {\hat{R}}}_i)$,  Eq.~(\ref{eq:t7}), and
\begin{eqnarray}
W(L,L_a) &=& (-1)^{L+L_a}\sqrt{\frac{30(2L+1)L(2L-1)}{(2L+3)(L+1)}}
\left\{%
\begin{matrix}
  1 & 1 & 2 \\
  L & L & L_a \\
\end{matrix}%
\right\}\,.
 \label {eq:t27}
\end{eqnarray}
In the above,  the set of
energies and wave functions $\{E_n(L_a),|nL_aM_a\rangle\}$ correspond to an intermediate energy
spectrum allowed by the dipole selection rule, which can be obtained
by diagonalizing the Hamiltonian in a Hylleraas basis set of given
symmetry $L_a$. In particular, for the case of $L=0$,
\begin{eqnarray}
\alpha_1 &=& \alpha_1(P)\,, \label{eq:t28}\\
\alpha_1^{(T)}&=& 0\,;
 \label {eq:t29}
\end{eqnarray}
for $L=1$,
\begin{eqnarray}
\alpha_1 &=& \alpha_1(S)+\alpha_1(P)+\alpha_1(D)\,,  \label {eq:t30}\\
\alpha_1^{(T)} &=&
-\alpha_1(S)+\frac{1}{2}\alpha_1(P)-\frac{1}{10}\alpha_1(D)\,;
 \label {eq:t31}
\end{eqnarray}
and for $L=2$,
\begin{eqnarray}
\alpha_1 &=& \alpha_1(P)+\alpha_1(D)+\alpha_1(F)\,, \label {eq:t32}\\
\alpha_1^{(T)} &=&
-\alpha_1(P)+\alpha_1(D)-\frac{2}{7}\alpha_1(F)\,. \label {eq:t33}
\end{eqnarray}
In  Eqs.~(\ref{eq:t30}) and (\ref{eq:t31}),   $\alpha_1(P)$ is the contribution from the even-parity
configuration $(pp')P$. In Eqs.~(\ref{eq:t32}) and
(\ref{eq:t33}), $\alpha_1(D)$ is from the odd-parity configuration $(pd)D$.

The fourth-order energy shift can be written in the form
\begin{eqnarray}
\Delta E_4 &=&
-\frac{{\cal{E}}^4}{24}\bigg[\gamma_0+\gamma_2\,g_2(L,M)
+\gamma_4\,g_4(L,M)\bigg]\,,
 \label {eq:t34}
\end{eqnarray}
where $g_4(L,M)$ is given by Eq.~(\ref{eq:a38}) in the Appendix. In
Eq.~(\ref{eq:t34}), $\gamma_0$ is the scalar second hyperpolarizability,
and $\gamma_2$ and $\gamma_4$ are the tensor second hyperpolarizabilities,
which can be written as
\begin{eqnarray}
\gamma_{0}&=&(-1)^{2L}\frac{128\pi^{2}}{3}\frac{1}{\sqrt{2L+1}}\sum_{L_{a}L_{b}L_{c}}
\mathcal {G}_{0}(L,L_{a},L_{b},L_{c})\mathcal{T}(L_{a},L_{b},L_{c}),
\label {eq:t35}
\end{eqnarray}
\begin{eqnarray}
\gamma_{2}&=&(-1)^{2L}\frac{128\pi^{2}}{3}\sqrt{\frac{L(2L-1)}{(2L+3)(L+1)(2L+1)}}
\sum_{L_{a}L_{b}L_{c}}
\mathcal{G}_{2}(L,L_{a},L_{b},L_{c})\mathcal{T}(L_{a},L_{b},L_{c}),
\label {eq:t36}
\end{eqnarray}
\begin{eqnarray}
\gamma_{4}&=&(-1)^{2L}\frac{128\pi^{2}}{3}
\sqrt{\frac{L(2L-1)(L-1)(2L-3)} {(2L+5)(L+2)(2L+3)(L+1)(2L+1)}}
\sum_{L_{a}L_{b}L_{c}}
\mathcal{G}_{4}(L,L_{a},L_{b},L_{c})\mathcal{T}(L_{a},L_{b},L_{c}),
\label {eq:t37}
\end{eqnarray}
where $\mathcal{T}(L_a,L_b,L_c)$ and $\mathcal{G}_{\Lambda}(L,L_{a},L_{b},L_{c})$,
 respectively,
are defined in the Appendix by Eqs.~(\ref{eq:a35}) and (\ref{eq:a42}). In particular,
for the case  $L=0$ we
only need to consider
\begin{eqnarray}
\gamma_0 &=&
\dfrac{128\pi^2}{3}\left[\frac{1}{9}\mathcal{T}(1,0,1)+\frac{2}{45}\mathcal{T}(1,2,1)\right]\,.
 \label {eq:t38}
\end{eqnarray}
[Note that the case $L_b=1$ does not enter in Eq.~(\ref{eq:t38}).  The
first 3-$j$ symbol in (\ref{eq:a42}) requires $K_1=0$ or $2$ and since $L=0$ here, the first 6-$j$
symbol requires $L_b=K_1$; therefore,  $L_b=0$ or $2$.]
For the case $L=1$, we have
\begin{eqnarray}
\gamma_{0}&=&\frac{128\pi^{2}}{3}\bigg[\frac{1}{27}\mathcal{T}(0,1,0)+\frac{2}{135}\mathcal{T}(0,1,2)
+\frac{1}{54}\mathcal{T}(1,1,1)-\frac{1}{90}\mathcal{T}(1,1,2)-\frac{1}{90}\mathcal{T}(1,2,1)\nonumber\\
&-&\frac{\sqrt{5}}{450}\mathcal{T}(1,2,2)+\frac{2}{135}\mathcal{T}(2,1,0)-\frac{1}{90}\mathcal{T}(2,1,1)
+\frac{17}{1350}\mathcal{T}(2,1,2)-
\frac{\sqrt{5}}{450}\mathcal{T}(2,2,1)\nonumber\\
&-&\frac{1}{450}\mathcal{T}(2,2,2)+\frac{2}{225}\mathcal{T}(2,3,2)\bigg]\,,\label{eq:t39}\\
\gamma_{2}&=&\frac{128\pi^{2}}{3}\bigg[-\frac{1}{27}\mathcal{T}(0,1,0)-\frac{2}{135}\mathcal{T}(0,1,2)
+\frac{1}{108}\mathcal{T}(1,1,1)-\frac{1}{180}\mathcal{T}(1,1,2)-\frac{1}{180}\mathcal{T}(1,2,1)\nonumber\\
&-&\frac{\sqrt{5}}{900}\mathcal{T}(1,2,2)-\frac{2}{135}\mathcal{T}(2,1,0)-\frac{1}{180}\mathcal{T}(2,1,1)
-\frac{7}{2700}\mathcal{T}(2,1,2)-
\frac{\sqrt{5}}{900}\mathcal{T}(2,2,1)\nonumber\\
&-&\frac{1}{900}\mathcal{T}(2,2,2)-\frac{2}{1575}\mathcal{T}(2,3,2)\bigg]\,
\label{eq:t40}
\end{eqnarray}
and for $L=2$, we  have
\begin{eqnarray}
\gamma_{0}&=&\frac{128\pi^{2}}{3}\cdot\frac{1}{15750}\bigg\{\bigg[140\,\mathcal{T}(1,0,1)
+119\,\mathcal{T}(1,2,1)+84\,\mathcal{T}(1,2,3)
+84\,\mathcal{T}(3,2,1)+74\,\mathcal{T}(3,2,3)+60\,\mathcal{T}(3,4,3)\bigg]\nonumber\\
&-&\bigg[105\,\mathcal{T}(1,1,1)+21\sqrt{5}\,\mathcal{T}(1,1,2)+21\,\mathcal{T}(1,2,2)
+21\sqrt{5}\,\mathcal{T}(2,1,1)+21\,\mathcal{T}(2,1,2)+21\,\mathcal{T}(2,2,1)-119\,\mathcal{T}(2,2,2)\nonumber\\
&+&56\,\mathcal{T}(2,2,3)+56\,\mathcal{T}(2,3,2)+4\sqrt{70}\,\mathcal{T}(2,3,3)+56\,\mathcal{T}(3,2,2)
+4\sqrt{70}\,\mathcal{T}(3,3,2)+20\,\mathcal{T}(3,3,3)\bigg]\bigg\}\,,\label{eq:t41}\\
\gamma_{2}&=&\frac{128\pi^{2}}{3}\cdot\frac{1}{154350}\bigg\{-\bigg
[1960\,\mathcal{T}(1,0,1)+1225\,\mathcal{T}(1,2,1)+840\,\mathcal{T}(1,2,3)
+840\,\mathcal{T}(3,2,1)+380\,\mathcal{T}(3,2,3)+240\,\mathcal{T}(3,4,3)\bigg]\nonumber\\
&+&\bigg[735\,\mathcal{T}(1,1,1)+147\sqrt{5}\,\mathcal{T}(1,1,2)+147\,\mathcal{T}(1,2,2)
+147\sqrt{5}\,\mathcal{T}(2,1,1)+147\,\mathcal{T}(2,1,2)+147\,\mathcal{T}(2,2,1)+1519\,\mathcal{T}(2,2,2)\nonumber\\
&-&448\,\mathcal{T}(2,2,3)-448\,\mathcal{T}(2,3,2)-32\sqrt{70}\,\mathcal{T}(2,3,3)-448\,\mathcal{T}(3,2,2)
-32\sqrt{70}\,\mathcal{T}(3,3,2)-160\,\mathcal{T}(3,3,3)\bigg]\bigg\}\,,\label{eq:t42}\\
\gamma_{4}&=&\frac{128\pi^{2}}{3}\cdot\frac{1}{128625}\bigg\{\bigg[490\,\mathcal{T}(1,0,1)
+49\,\mathcal{T}(1,2,1)+14\,\mathcal{T}(1,2,3)
+14\,\mathcal{T}(3,2,1)+4\,\mathcal{T}(3,2,3)+\frac{5}{3}\,\mathcal{T}(3,4,3)\bigg]\nonumber\\
&+&\bigg[245\,\mathcal{T}(1,1,1)+49\sqrt{5}\,\mathcal{T}(1,1,2)+49\,\mathcal{T}(1,2,2)
+49\sqrt{5}\,\mathcal{T}(2,1,1)+49\,\mathcal{T}(2,1,2)+49\,\mathcal{T}(2,2,1)+49\,\mathcal{T}(2,2,2)\nonumber\\
&+&14\,\mathcal{T}(2,2,3)+14\,\mathcal{T}(2,3,2)+\sqrt{70}\,\mathcal{T}(2,3,3)+14\,\mathcal{T}(3,2,2)
+\sqrt{70}\,\mathcal{T}(3,3,2)+5\,\mathcal{T}(3,3,3)\bigg]\bigg\}\,.\label{eq:t43}
\end{eqnarray}
In each of Eqs.~(\ref{eq:t41})-(\ref{eq:t43}), the terms in the first set of square
brackets only involve the intermediate states of natural parities,
which make the dominant contributions to the hyperpolarizabilities,
while
the terms in the second set of square brackets involve the
intermediate states of unnatural parities, which make subordinate
contributions. For example, the term $\mathcal{T}(1,0,1)$ involves
the intermediate states  $1s^2np\,^2\!P^{\rm o}$,
$1s^2ns\,^2\!S^{\rm e}$, and $1s^2np\,^2\!P^{\rm o}$, which are all
natural-parity states, while  the term $\mathcal{T}(1,1,1)$
involves the electronic configurations $1s^2np\,^2\!P^{\rm o}$,
$1snpn'p\,^2\!P^{\rm e}$, and $1s^2np\,^2\!P^{\rm o}$, where two
states are natural-parity states and one unnatural-parity.

The scalar dipole polarizability defined in Eq.~(\ref{eq:t24}) can
be generalized to the $2^\ell$-pole polarizability $\alpha_\ell$
\begin{eqnarray}
\alpha_\ell &=& \sum_{L_a} \alpha_\ell(L_a)\,, \label {eq:t44}
\end{eqnarray}
where
\begin{eqnarray}
\alpha_\ell(L_a) &=& \frac{8\pi}{(2\ell+1)^2(2L+1)}\sum_n
\frac{|\langle n_0 L \|T_\ell\| nL_a\rangle
|^2}{E_n(L_a)-E_{n_0}(L)}\,,
 \label {eq:t45}
\end{eqnarray}
and $T_\ell$ is the $2^\ell$-pole transition operator given by
Eq.~(\ref{eq:t7}).

\subsection{Coefficients for long-range interactions between two atoms}\label {dispersion}

First, let us consider the simplest case where both a and b are  Li atoms in
their ground states~\cite{yan}. At large separations $R$, using
second-order perturbation theory, $V_{ab}$ can be expressed as a series
in inverse powers of $R$,
\begin{eqnarray}
V_{ab}=-\frac{C_{6}}{R^{6}}-\frac{C_{8}}{R^{8}}-\frac{C_{10}}{R^{10}}-\cdots\,,
 \label {eq:t46}
\end{eqnarray}
where
\begin{eqnarray}
C_{6}=\frac{3}{\pi}G_{ab}(1,1)\,, \label {eq:t47}
\end{eqnarray}
\begin{eqnarray}
C_{8}=\frac{15}{2\pi}G_{ab}(1,2)+\frac{15}{2\pi}G_{ab}(2,1)\,,
\label{eq:t48}
\end{eqnarray}
and
\begin{eqnarray}
C_{10}=\frac{14}{\pi}G_{ab}(1,3)+\frac{14}{\pi}G_{ab}(3,1)+\frac{35}{\pi}G_{ab}(2,2)\,.
\label {eq:t49}
\end{eqnarray}
Introducing the oscillator strength for the transition
$|n_0 L\rangle\rightarrow |n L'\rangle$
\begin{eqnarray}
\bar{f}_{n0}^{(\ell)}=\frac{8\pi}{(2\ell+1)^{2}(2L+1)}E_{n0}
|\langle n_0L\|T_\ell\|nL'\rangle|^{2}, \label{eq:t50}
\end{eqnarray}
where $E_{n0}=E_n(L')-E_{n_0}(L)$ is the corresponding transition
energy, $G_{ab}(\ell_a,\ell_b)$ can be written in the form
\begin{eqnarray}
G_{ab}(\ell_{a},\ell_{b})=\frac{\pi}{2}\sum_{nn'}\frac{\bar{f}_{n0}^{(\ell_{a})}
\bar{f}_{n'0}^{(\ell_{b})}}{E_{n0}^{a}E_{n' 0}^{b}(E_{n0}^{a}+E_{n'
0}^{b})}\,, \label{eq:t51}
\end{eqnarray}
where throughout $a$ and $b$, respectively, denote atom a and atom b.

Next, we consider two like lithium atoms a and b, where atom a is in
the ground state and atom b in an
excited state with  orbital angular momentum $L_b$ and
associated magnetic quantum number $M_b$. The zeroth order wave
function for the combined system ab can be written in the form~\cite{yan}:
\begin{eqnarray}
\Psi^{(0)}=\frac{1}{\sqrt{2}}[\Psi_a(\boldsymbol
{\sigma})\Psi_b(L_bM_b;\boldsymbol
{\rho})+\beta\Psi_a(\boldsymbol{\rho}) \Psi_b(L_bM_b;\boldsymbol
{\sigma})]\,,\label{eq:52}
\end{eqnarray}
where $\boldsymbol{\sigma}$ and $\boldsymbol{\rho}$ represent,
respectively, the set of all of  the internal
coordinates for atom a and atom b, and $\beta=\pm 1$
describes the symmetry of the system due to the exchange of two atoms.
According to the
perturbation theory, the first-order interaction energy is given by
\begin{eqnarray}
V^{(1)}(L_bM_b;\beta)=-\frac{C_{2L_b+1}^{M_b\beta}}{R^{2L_b+1}}\,,\label{eq:t53}
\end{eqnarray}
where
\begin{eqnarray}
C_{2L_b+1}^{M_b\beta}=\beta(-1)^{1+L_b+M_b}\frac{4\pi}{(2L_b+1)^2}\left(
\begin{array}{c}
2L_b\\
L_b+M_b
\end{array}\right) |\langle\Psi_a(\boldsymbol
{\sigma})\|T_{L_b}(\boldsymbol
{\sigma})\|\Psi_b(L_b;\boldsymbol {\sigma})\rangle|^2\,.
\label{eq:t54}
\end{eqnarray}
One can see from Eq.~(\ref{eq:t53}) that, for the Li($S$)-Li($P$) system,
the interaction energy is proportional to $R^{-3}$.
To get the next order energy, let the complete set for the intermediate states of the system be
\begin{eqnarray}
\{\chi_s(L_sM_s;\boldsymbol
{\sigma})\omega_t(L_tM_t;\boldsymbol{\rho})\}\,, \label{eq:t55}
\end{eqnarray}
with the energy eigenvalue $E_{st}^{(0)}=E_s^{(0)}+E_t^{(0)}$. According to the second-order perturbation theory, the second-order
energy is
\begin{eqnarray}
V^{(2)}=-\frac{C_6^{M_b}}{R^6}-\frac{C_8^{M_b}}{R^8}-\cdots,\label{eq:t56}
\end{eqnarray}
where
\begin{eqnarray}
C_6^{M_b}&=&\sum_{st}\frac{\Omega_6^{st}}{E_{st}^{(0)}-E^{(0)}}\,\label{eq:t57}\,, \\
C_8^{M_b}&=&\sum_{st}\frac{\Omega_8^{st}}{E_{st}^{(0)}-E^{(0)}}\,\label{eq:t58}\,,
\end{eqnarray}
and
the energy for
the unperturbed system is $E^{(0)}=E_a^{(0)}+E_b^{(0)}$.
Following \cite{yan}, one can obtain the following expressions for $\Omega_6^{st}$ and $\Omega_8^{st}$ that
are in agreement with the formulas in Ref.~\cite{ZhaYanVri06a}:
\begin{eqnarray}
\Omega_6^{st}=|\langle\Psi_a(\boldsymbol
{\sigma})\|T_1(\boldsymbol {\sigma})\|\chi_s(1;\boldsymbol
{\sigma})\rangle|^2 \sum_\lambda
G(1,1,1,\lambda,1,M_b)|\langle\Psi_b(1;\boldsymbol{\rho})\|T_1(\boldsymbol
{\rho})\|\omega_t(\lambda;\boldsymbol{\rho})\rangle|^2\,,\label{eq:t59}
\end{eqnarray}
\begin{eqnarray}
\Omega_8^{st}&=&2\,|\langle\Psi_a(\boldsymbol
{\sigma})\|T_1(\boldsymbol {\sigma})\|\chi_s(1;\boldsymbol
{\sigma})\rangle|^2\sum_\lambda
G(1,3,1,\lambda,1,M_b)\langle\Psi_b(1;\boldsymbol{\rho})\|T_1(\boldsymbol
{\rho})\|\omega_t(\lambda;\boldsymbol{\rho})\rangle\nonumber\\
&&\times \langle\Psi_b(1;\boldsymbol{\rho})\|T_3(\boldsymbol
{\rho})\|\omega_t(\lambda;\boldsymbol{\rho})\rangle\nonumber\\
&+& |\langle\Psi_a(\boldsymbol {\sigma})\|T_1(\boldsymbol
{\sigma})\|\chi_s(1;\boldsymbol
{\sigma})\rangle|^2\sum_\lambda
G(2,2,1,\lambda,1,M_b)|\langle\Psi_b(1;\boldsymbol{\rho})\|T_2(\boldsymbol
{\rho})\|\omega_t(\lambda;\boldsymbol{\rho})\rangle|^2\nonumber\\
&+& |\langle\Psi_a(\boldsymbol {\sigma})\|T_2(\boldsymbol
{\sigma})\|\chi_s(2;\boldsymbol
{\sigma})\rangle|^2\sum_\lambda
G(1,1,2,\lambda,1,M_b)|\langle\Psi_b(1;\boldsymbol{\rho})\|T_1(\boldsymbol
{\rho})\|\omega_t(\lambda;\boldsymbol{\rho})\rangle|^2\nonumber\\
&+& G_3(2,2,1,1,1,M_b)\langle\Psi_a(\boldsymbol {\sigma})\|T_1(\boldsymbol
{\sigma})\|\chi_s(1;\boldsymbol {\sigma})\rangle
\langle\Psi_a(\boldsymbol{\rho})\|T_1(\boldsymbol
{\rho})\|\omega_t(1;\boldsymbol{\rho})\rangle\nonumber\\
&&\times \langle\Psi_b(1;\boldsymbol{\rho})\|T_2(\boldsymbol
{\rho})\|\omega_t(1;\boldsymbol{\rho})\rangle \langle\Psi_b(1;\boldsymbol
{\sigma})\|T_2(\boldsymbol {\sigma})\|\chi_s(1;\boldsymbol
{\sigma})\rangle\nonumber\\
&+&2\,G_3(1,2,1,2,1,M_b)\langle\Psi_a(\boldsymbol {\sigma})\|T_1(\boldsymbol
{\sigma})\|\chi_s(1;\boldsymbol {\sigma})\rangle
\langle\Psi_a(\boldsymbol{\rho})\|T_2({\boldsymbol{\rho}})
\|\omega_t(2;\boldsymbol{\rho})\rangle\nonumber\\
&&\times \langle\Psi_b(1;\boldsymbol{\rho})\|T_1({\boldsymbol{\rho}}
)\|\omega_t(2;\boldsymbol{\rho})\rangle
\langle\Psi_b(1;\boldsymbol{\sigma})\|T_2(\boldsymbol{\sigma})
)\|\chi_s(1;\boldsymbol{\sigma})\rangle\nonumber\\
&+&G_3(1,1,2,2,1,M_b)\langle\Psi_a(\boldsymbol{\sigma})\|T_2(\boldsymbol{\sigma})
)\|\chi_s(2;\boldsymbol{\sigma})\rangle
\langle\Psi_a(\boldsymbol{\rho})\|T_2(\boldsymbol{\rho}
)\|\omega_t(2;\boldsymbol{\rho})\rangle\nonumber\\
&&\times \langle\Psi_b(1;\boldsymbol{\rho})\|T_1(\boldsymbol{\rho}
)\|\omega_t(2;\boldsymbol{\rho})\rangle
\langle\Psi_b(1;\boldsymbol{\sigma})\|T_1(\boldsymbol{\sigma}
)\|\chi_s(2;\boldsymbol{\sigma})\rangle\,,\label{eq:t60}
\end{eqnarray}
with
\begin{eqnarray}
G(L,L',L_s,L_t,L_b,M_b)&=&(-1)^{L+L'}\frac{(4\pi)^2}{(2L_s+1)^2}(L,L')^{-1/2}\sum_{M_sM_t}K_{L_sL}^{-M_s}
K_{L_sL'}^{-M_s}\nonumber\\
&&\left(\begin{array}{ccc} L_b&L&L_t\\
-M_b&M_s&M_t
\end{array}
\right)
\left(\begin{array}{ccc} L_b&L'&L_t\\
-M_b&M_s&M_t
\end{array}
\right)\,,\label{eq:t61}
\end{eqnarray}
and
\begin{eqnarray}
G_3(L,L',L_s,L_t,L_b,M_b)&=&(-1)^{L+L_s}
\frac{(4\pi)^2}{(2L_s+1)(2L_t+1)}(L,L')^{-1/2}\sum_{M_sM_t}(-1)^{M_s+M_t}K_{L_sL}^{-M_s}
K_{L'L_t}^{M_t}\nonumber\\
&&\left(\begin{array}{ccc} L_b&L&L_t\\
-M_b&M_s&M_t
\end{array}
\right)
\left(\begin{array}{ccc} L_b&L'&L_s\\
-M_b&M_t&M_s
\end{array}
\right)\,.\label{eq:t62}
\end{eqnarray}
In Eqs.~(\ref{eq:t61}) and (\ref{eq:t62}), the coefficient
$K_{\ell L}^\mu$ is
\begin{eqnarray}
K_{\ell L}^\mu=\left [ \left(
\begin{array}{c}
\ell+L\\
\ell+\mu
\end{array}
\right)
\left(
\begin{array}{c}
\ell+L\\
L+\mu \end{array} \right) \right]^{1/2}\,,
\end{eqnarray}
and $(\ell,L,\cdots)=(2\ell+1)(2L+1)\cdots$.

\section{Results and discussions}\label {result}

In this section, we present the results of the
calculations of
the static electric dipole, quadrupole, and octupole polarizabilities,
the second hyperpolarizabilities, and the resonance and dispersion long-range
coefficients for the pairs of atoms using the
wave functions obtained as described in Sec.~\ref{basis}.
In addition, we give some calculated oscillator strengths and sum rules
that might be useful.

\subsection{$2\,^2\!S$ state: Polarizabilities
and hyperpolarizabilities of atoms and dispersion coefficients between two atoms}

Table~\ref{g5} presents a convergence study for the calculations of
the scalar dipole polarizability $\alpha_1$ of lithium
with infinite nuclear mass, $^\infty$Li, in the
ground state. In the table, $N_0$ and $N_P$, respectively, are
the sizes of the basis sets for the ground state and  for the intermediate
states of symmetry $P$. The extrapolation, obtained  by assuming that the
ratio between two successive differences in $\alpha_1$ stays
constant as the sizes of the basis sets become infinitely large,
yields the
value 164.112(1). This is  in perfect agreement with the
value 164.111(2) of  Ref.~\cite{yan},  based on
calculations up to much smaller
values $N_0=919$ and $N_P=1846$,
confirming the efficacy of the extrapolation method used in that work.
For $^6$Li and $^7$Li, a similar
convergence pattern exists.

Calculations on the  hyperpolarizability $\gamma_0$, on the other hand, require much larger basis
set sizes to achieve accuracies even approaching that achieved
for the polarizabilities.
In our approach, there is
a partial cancellation of significant figures between the sum of the two terms
$\textstyle{\frac{1}{9}}\,\mathcal{T}(1,0,1)$
and $\textstyle{\frac{2}{45}}\,\mathcal{T}(1,2,1)$
in
Eq.~(\ref{eq:t38}), even though the individual terms are converged
to about 4 significant figures.
Table~\ref{g7} presents the convergence study for calculations of
the hyperpolarizability $\gamma_0$ of $^\infty$Li in the ground
state.
At the  the largest sizes
of basis sets in Table~\ref{g7}, $\textstyle{\frac{1}{9}}\,\mathcal{T}(1,0,1) =
-3463.861$ and
$\textstyle{\frac{2}{45}}\,\mathcal{T}(1,2,1)=3471.078$, resulting
in a loss of about two significant figures when added.

Table~\ref{g2} summarizes the final values of the polarizabilities
$\alpha_1$, $\alpha_2$, and $\alpha_3$,
$\gamma_0$, and the dispersion coefficients $C_6$, $C_8$, and
$C_{10}$ for $^\infty$Li, $^7$Li, and $^6$Li in their ground states
$2\,^2\!S$.
The dispersion coefficients were calculated similarly to those
calculated in Ref.~\cite{yan}, but with the treatment
of finite nuclear mass.
In order to  maintain numerical stability,
especially in diagonalizing the Hamiltonian for large basis sets,
all the calculations were performed using the multiple precision
arithmetic QD developed by Bailey {\it et al.}~\cite{HidLiBai-QD}, which
has 64 decimal digits.
The finite nuclear mass effect is most significant for $\gamma_0$, as
shown in Table~\ref{g2}.
Fitting the results, we find, roughly, that
$\gamma_0\sim 3060 [ 1- 1000 (1/m_0)]$,
where $m_0$ is the nuclear mass
and where the coefficient of $1/m_0$ is about 1000 times larger
than it is for, say, $\alpha_1$.
The effect is numerically significant since, as discussed above, our convergence
studies show that at least two digits of the hyperpolarizabilities
are accurate.
It would be interesting to explore this effect for other systems.

In Table~\ref{g6} we compare our results
to some of those from the  literature for the polarizabilities
$\alpha_1$, $\alpha_2$, and $\alpha_3$ and for the dispersion
coefficients $C_6$, $C_8$, and $C_{10}$ of ground state Li atoms.
As most of
the previously published works have been devoted to $^\infty$Li,
we list our infinite nuclear mass calculations in Table~\ref{g6}.
The most accurate Hylleraas-type calculation prior to the present work
is that of Ref.~\cite{yan},  which was obtained
using a basis size up to  1846 using methods similar
to the present work,
extrapolated as discussed previously.
An extensive tabulation of over forty results
for $\alpha_1$ covering much of the published work from 1959
to 1996 can be found in the review article by King~\cite{Kin97}
and
another tabulation is given in Ref.~\cite{yan}.
Tabulations including $\alpha_2$ and $\alpha_3$ are given
in Refs.~\cite{yan} and \cite{CheWan04}.
In Table~\ref{g6} we  tabulate some theoretical results
from between 1996 and the present.
The exponentially
correlated Gaussian-basis set calculations of Komasa~\cite{Kom01}
for $\alpha_1$
are in excellent agreement with the present work
and they were carried out with a much smaller basis size.
There is a slight discrepancy, however, for $\alpha_2$,
but Ref.~\cite{Kom01} does not provide computational uncertainties.
Both
$\alpha_1$ and $\alpha_2$ were calculated  using
semi-empirical model potential-based methods by Cohen and
Themelis~\cite{CohThe05}
and by Zhang {\it et
al.}~\cite{ZhaMitBro07}.
Cohen and
Themelis~\cite{CohThe05}
used a method
dubbed Rydberg-Klein-Rees quantum defect theory (RKR-QDT),
which utilized  experimental
energy levels and it contained some adjustable parameters
fixed using the experimentally determined $2p$ state lifetime.
The results of Zhang \textit{et
al.}~\cite{ZhaMitBro07} were obtained in the framework of a frozen core
Hamiltonian with a semi-empirical polarization potential.
Of the two semi-empirical approaches, the RKR-QDT results
are in much better agreement with the present work;
the results of Zhang \textit{et al.} for $\alpha_3$ vary
from the present work by the same percentage as do
their calculations of $\alpha_1$ and $\alpha_2$.
Chen and Wang~\cite{CheWan04}
evaluated $\alpha_2$ and $\alpha_3$ for the ground states of
lithium-like ions using the full core plus correlation method.
The present values lie just outside the lower limit of Chen and Wang's
error bars, but Ref.~\cite{CheWan04} does
not reveal how the uncertainties were obtained.

We can also compare our results to recent relativistic calculations.
For $\alpha_1$ generally the effect is to reduce the
non-relativistic value by a term of order $(1/137.037)^2$.
Derevianko {\it et al.}~\cite{DerBabDal01} and Porsev and
Derevianko~\cite{PorDer03} calculated $\alpha_1$, $\alpha_2$, and
$\alpha_3$, as well as $C_6$, $C_8$, and $C_{10}$ for the ground
state of lithium using relativistic many-body perturbation theory
(MBPT). Sahoo {\it et al.}~\cite{Sah07,WanSahTim08} performed {\it
ab initio} relativistic coupled-cluster calculations on the dipole
and quadrupole polarizabilities, and the dispersion coefficients
$C_6$ and $C_8$ of Li. Very recently, Johnson {\it et
al.}~\cite{JohSafDer08} also calculated the polarizabilities of
$^7$Li by applying  relativistic MPBT. For $\alpha_1$, the results
of Refs.~\cite{DerBabDal01} and \cite{JohSafDer08}, obtained using
relativistic MPBT are  smaller than our results by the expected
factor. The results of Refs.~\cite{Sah07,WanSahTim08} are
significantly lower, perhaps due to neglected higher-order
effects~\cite{WanSahTim08}, see Table~\ref{g6}.

For $C_6$, compared to the previous value~\cite{yan}, the
uncertainty in the present result has been reduced by a factor of 3.
The semi-empirical calculations of Zhang~\textit{et
al.}~\cite{ZhaMitBro07} and the relativistic calculations of
Refs.~\cite{DerBabDal01} and \cite{PorDer03} are in good agreement
with the present results, though Zhang~\textit{et al.} slightly
overestimate the coefficients, as shown in Table~\ref{g6}.

The most precise measurement of $\alpha_1$ was that of  Miffre {\it et
al.}~\cite{MifJacMuc06}  obtained with 0.66\% uncertainty  using atomic
interferometry. It is a factor of
three more precise than the earlier measurement of Molof {\it et al.}~\cite{MolSchMil74}.
These values are included in Table~\ref{g6}.

Currently, however, the experimental accuracy has not reached the stage where
finite nuclear mass and relativistic effects in $\alpha_1$ can be
tested stringently. One can see from Table~\ref{g2} that the finite
nuclear mass correction to $\alpha_1$ for  $^7$Li is 0.049~a.u., of
which 0.038~a.u. comes simply from the mass scaling of the Bohr
radius. The remaining part
 0.011~a.u. is due to the mass polarization
terms $(-1/m_0)\,\nabla_i\cdot\nabla_j$ in the Hamiltonian
Eq.~(\ref{eq:t6}), where $m_0$ is the mass of the $^7$Li nucleus.
Furthermore, although the relativistic effect has not been evaluated
rigorously, it can be, however, estimated to be $-0.06$~a.u. based on a
relativistic coupled-cluster
approach~\cite{Sch06,LimPerSet99,ThaLup05}. As for the QED
effect, Pachucki and Sapirstein~\cite{PacSap00} performed a
relativistic and QED calculation on the dipole polarizability of
helium and found that the QED correction is about a factor of 2.5
smaller than the relativistic correction and is opposite in sign. If
we take this reduction factor for the case of lithium, the QED
correction is thus estimated to be 0.02~a.u. Hence, the finite
nuclear mass, relativistic, and QED corrections cancel out almost
entirely, just as in the case of helium. However, a definitive
conclusion regarding this cancellation can not be drawn until the
calculation of Pachucki and Sapirstein can be extended to the case
of Li.

As discussed above, we have found that the hyperpolarizability
is extraordinarily sensitive to the finite nuclear mass.
In Table~\ref{hyper-results} we compare our calculated
value of $\gamma_0$  with
some of the published results, all for $^\infty$Li.
A more comprehensive table of earlier work is given in Ref.~\cite{Kin97}.
Pipin and Bishop~\cite{PipBis92} calculated $\gamma_0$ by
applying the combined configuration-interaction-Hylleraas method.
Their result, with one significant figure,
is in good agreement with ours. Kassimi and
Thakkar~\cite{KasTha94,ThaLup06} used the coupled-cluster
approach, where the reported uncertainty of $10\%$ in $\gamma_0$ was
later re-adjusted to $3\%$~\cite{ThaLup06}.
Laughlin~\cite{Lau95} performed a semi-empirical one-electron model potential
calculation and found that the final result
for $\gamma_0$ was highly sensitive to the data used,
particularly the value of $\alpha_1$. The value from
Jaszu\`{n}ski and Rizzo~\cite{JasRiz96} was obtained using a
series of multi-configuration SCF wave functions. Finally, Cohen and
Themelis~\cite{CohThe05} computed $\gamma_0$ using RKR-QDT.
Comparing to our result, the
level of  accuracy they achieved is about 10\%.
The RKR-QDT calculation was sensitive to the potential adopted,
as discussed in Ref.~\cite{CohThe05}.
The present methodology has the advantage that no adjustment is required.
Once the convergence pattern is established, the extrapolated
value should be reliable.

\subsection{$2\,^2\!P$ state: Polarizabilities and hyperpolarizabilities}

Table~\ref{g8} shows the  convergence of $\alpha_1$ and of
$\alpha_1^{(T)}$ calculated for $^{\infty}$Li in the $2\,^2\!P$
state, where $N_S$, $N_{(pp')P}$, and $N_D$ are the sizes of bases
for the intermediate states of symmetries $S$, $P$, and $D$
respectively, and $(pp')P$ stands for the main configuration of two
$p$-electrons coupled to form a total angular momentum of $P$. Since
the contribution from the $(pp')P$ configuration is well converged at
$N_{(pp')P}=3413$, we did not increase $N_{(pp')P}$ any further.

Table~\ref{g3} summarizes the final values of the
scalar polarizabilities $\alpha_1$, $\alpha_2$, and $\alpha_3$, the
tensor dipole polarizability $\alpha_1^{(T)}$, the scalar
hyperpolarizability $\gamma_0$, and the tensor hyperpolarizability
$\gamma_2$ for $^\infty$Li, $^7$Li, and $^6$Li in their
$2\,^2\!P$ states.

Table~\ref{g9} is a comparison of the scalar and tensor dipole polarizabilities and
hyperpolarizabilities $\alpha_1$, $\alpha_1^{(T)}$, $\gamma_0$, and
$\gamma_2$ for the $2\,^2\!P$ state of $^\infty$Li. In general, our
calculations provide significantly more accurate results,
particularly, for $\alpha_1^{(T)}$.
A more extensive tabulation of previous results is given in Ref.~\cite{CohThe05}.

It should be mentioned that the
intermediate configuration of symmetry $(pp')P$, which contributes
to $\alpha_1$ at the level of 0.05\%, was not included in the
CI-Hylleraas calculation of Pipin and Bishop~\cite{PipBis93}.
The relativistic results for $\alpha_1$
by Wansbeek {\it et al.}~\cite{WanSahTim08}
and by Johnson {\it et al.}~\cite{JohSafDer08} are the $(2J+1)$-weighted sums between
the $J=1/2$ and $J=3/2$ sub-levels.
The results of Johnson  {\it et al.} are close to
ours. In contrast, the results of
Wansbeek {\it et al.}~\cite{WanSahTim08} deviate significantly from ours;
for example, the
value of $\alpha_1^{(T)}$ reported by
Wansbeek {\it et al.}~\cite{WanSahTim08} is a factor of 3.6 larger than our calculation.

The uncertainties in the  experimental values for
$\alpha_1$ and $\alpha_1^{(T)}$ obtained by Hunter {\it et al.}~\cite{HunKraBer91}
and by Windholz \textit{et al.}~\cite{WinMusZer92} are too large to
reveal finite nuclear mass and relativistic
effects.
However, the nonrelativistic calculations of Refs.~\cite{PipBis93}, \cite{ZhaMitBro07},
and the present
work, as well as
the relativistic calculations of Ref.~\cite{JohSafDer08},  agree with the experiment of Windholz~\textit{et al.},
though
in contrast, as was also observed by Johnson~\textit{et al.}, the values
obtained using RKR-QDT by Cohen and Themelis~\cite{CohThe05} \textit{disagree}
with the experiment.

For the  $\alpha_2$ and $\alpha_3$   of the
$2\,^2\!P$ state, the model potential results of Zhang {\it et
al.}~\cite{ZhaMitBro07} are, respectively, 4104.9 and $3.2135\times 10^5$,
which are slightly larger than our results, given in Table~\ref{g3}.

There are no measurements of $\gamma_0$ and $\gamma_2$
though our calculated values are in almost perfect agreement with,
though substantially more accurate than, the semi-empirical
results of Cohen and Themelis~\cite{CohThe05} and
the early calculation of Themelis and Nicolaides~\cite{TheNic92},
obtained by fitting electric-field induced energy shifts calculated
using Hartree-Fock wave functions.

\subsection{Long-range interactions between a $2\,^2\!S$ atom and a $2\,^2\!P$ atom}

Table~\ref{g3a} lists the final values of  $C_3$, $C_6$, and
$C_{8}$ for the system $2\,^2\!S$--$2\,^2\!P$ between two like-atoms
$^\infty$Li, $^7$Li, and $^6$Li with all possible symmetries.

Table~\ref{g11} contains comparisons for $C_3$, $C_6$, and $C_8$
for the system $^\infty$Li($2\,^2S$)--$^\infty$Li($2\,^2\!P$)
with some published results, including the Hylleraas-type calculations
of Yan {\it et al.}~\cite{yan}, the model potential approach of Marinescu and
Dalgarno~\cite{MarDal95}, and the semi-empirical
model potential of Zhang {\it et al.}~\cite{ZhaMitBro07}.
Our \textit{ab initio} results confirm the more accurate
semi-empirical results of Zhang {\it et al.}~\cite{ZhaMitBro07},
though their results are systematically slightly larger in magnitude than ours.
It is also evident that
the present results
have substantially improved  the precision of  $C_8$.

\subsection{$3\,^2\!D$ state: Polarizabilities and hyperpolarizabilities}

Tables~\ref{g10a} and \ref{g10b}, respectively,  list  convergence studies
for the dipole polarizabilities $\alpha_1$ and $\alpha_1^{(T)}$ and for
the hyperpolarizabilities $\gamma_0$, $\gamma_2$,
and $\gamma_4$ of $^{\infty}$Li in $3\,^2\!D$.

Table~\ref{g4} is the summary of all the
values for the polarizabilities and the hyperpolarizabilities of
$^\infty$Li, $^7$Li, and $^6$Li in their $3\,^2\!D$ states.

Table~\ref{g10} presents comparisons of our calculated
scalar and tensor dipole polarizabilities
of $^\infty$Li
with the existing theoretical and experimental values for the $3\,^2\!D$ state.
The relative signs and magnitudes (though of limited accuracy)  were
correctly predicted by
Themelis and Nicolaides~\cite{TheNic95} using an empirically
modeled dipole moment operator
and fitting
to field-dependent energy shifts.
For $\alpha_1$,
the percentage difference between the semi-empirical model potential result of Magnier
and Aubert-Fr\'{e}con~\cite{MagAubFre02} and ours is about 7\%.
The values of Ashby and van Wijngaarden~\cite{AshvanWij03}, obtained using the
semi-empirical Coulomb
approximation method and the values of Zhang {\it et
al.}~\cite{ZhaMitBro07} obtained in the framework of a frozen core
Hamiltonian with a semi-empirical model potential, are very
close to each other, but they differ from our results at the
levels of, respectively, 0.9\% and 0.8\%.
The
experimental results of Ashby {\it et al.}~\cite{AshClavan03}
and the relativistic calculations of Wansbeek \textit{et al.}~\cite{WanSahTim08} are
expressed in the $LSJ$ coupling scheme, which may be converted into
the $LS$ coupling by averaging
over the fine structure using a $(2J+1)$-weighted sum~\cite{ZhaMitBro07}. One can
see that our result for $\alpha_1$ disagrees with the experiment at the level of
1\%. Similarly, for $\alpha_1^{(T)}$, the discrepancy
is at the level of 0.5\%. It should be pointed out that, at this level of accuracy, the
finite nuclear mass and relativistic effects do  not account for
the discrepancy.
The present calculation and that of Wansbeek \textit{et al.}~\cite{WanSahTim08}
are \textit{ab initio}. Their results, which
may suffer from an incomplete
treatment of correlation effects, differ drastically from our own.

\subsection{Some oscillator strengths and sum rules}

As by-products in calculating the dipole polarizabilities, we have
obtained the oscillator strengths of $^\infty$Li, $^7$Li, and $^6$Li
for the transitions of $2\,^2\!S-2\,^2\!P$, $2\,^2\!P-3\,^2\!D$, and
$3\,^2\!D-4\,^2\!F$, listed in Table~\ref{g0}. For the case of
$^\infty$Li, a comparison with some previous results is presented in
Table~\ref{g1}, including the multiconfiguration Hartree-Fock of
Godefroid {\it et al.}~\cite{GodFroFis01}, the semi-empirical model
potential of Zhang {\it et al.}~\cite{ZhaMitBro07}, and the
relativistic many-body approach of Johnson~\cite{JohSafDer08}. The
nonlinear variational parameters in our calculation are optimized
only for the lowest energy eigenstate of given symmetry and thus
individual  oscillator strengths for excited state transitions may
not necessarily be of better accuracy than other dedicated
calculations. Nevertheless, the present calculations have slightly
improved the previous values~\cite{YanDra95b} for the
$2\,^2\!S-2\,^2\!P$ and $2\,^2\!P-3\,^2\!D$ transitions, which were
obtained variationally using smaller Hylleraas basis sets up to
about 3500 terms. The most accurate results for the
$2\,^2\!S-2\,^2\!P$ oscillator strength was obtained by Yan
\textit{et al.}~\cite{YanTamDra98} using variational trial functions
that contained the core and valence electron wave functions in the
zeroth order. [Note that for $^7$Li and $^6$Li, the present
definition for the oscillator strength Eq.~(\ref{eq:t50}) differs
from the one adopted in Refs.~\cite{YanDra95b,YanTamDra98} by a
factor of $1+ (3\,/m_0)$, where $m_0$ is the nuclear mass.]

Finally, we have calculated the oscillator strength sum
rule~\cite{KouMea77} $S(-3)\equiv 2 \beta_1$, where  $\beta_1$ is
given in terms of the quantities defined in Sec.~\ref{theory}, in a
manner similar to the definition of $\alpha_l$ in
Eqs.~(\ref{eq:t44}) and (\ref{eq:t45}), as
\begin{eqnarray}
\beta_1 &=& \sum_{L_a} \beta_1(L_a),
\end{eqnarray}
where
\begin{eqnarray}
\beta_1(L_a) &=& \frac{4\pi}{9(2L+1)}\sum_n \frac{|\langle n_0 L
\|T_1\| nL_a\rangle |^2}{[E_n(L_a)-E_{n_0}(L)]^2} .
\end{eqnarray}

The results for $\beta_1$ are given in Table~\ref{sum-rule} for
$^\infty$Li, $^7$Li, and $^6$Li in the $2\,^2\!S$, $2\,^2\!P$, and
$3\,^2\!D$ states. The result for $^\infty$Li in the $2\,^2\!S$
state is in agreement with the value $\beta_1=1197$ given by Pipin
and Bishop~\cite{PipBis92}.

Another sum rule $S(-1)$ can be related to the interaction  potential between an atom and
a perfectly conducting wall, see~Ref.~\cite{YanDaBab97}. A highly accurate value
obtained~\cite{YanDaBab97} using matrix elements from Ref.~\cite{YanDra95a}
is $12.144\,004\,08(24)$ and the present work does not give an improvement.
The value from Bishop and Pipin~\cite{PipBis92} is 12.13.

\section{conclusions}\label {conclusions}

In this paper, the nonrelativistic polarizabilities and
hyperpolarizabilities have been calculated \textit{ab initio} in a
unified manner for the $2\,^2\!S$, $2\,^2\!P$, and $3\,^2\!D$ states
of lithium using fully-correlated Hylleraas basis sets. The
dispersion coefficients for Li($2\,^2S$)--Li($2\,^2\!S$) and
Li($2\,^2S$)--Li($2\,^2\!P$) have also been evaluated. Furthermore,
the finite nuclear mass effects on these properties have been
studied for Li laying the foundation for future work such as
investigating relativistic and QED effects on the polarizabilities,
especially on the dipole polarizability of the ground state of
lithium, using the approach of Pachucki and
Sapirstein~\cite{PacSap00}, following the progress on
He~\cite{PacSap00,CenSzaJez01,LacJezSca04,SchGavMay07}.

Our results can also be used as a benchmark for other methods that
may be developed in future research. For the dipole polarizabilities
of lithium in $3\,^2\!D$ state, an improved measurement would be
important in resolving the existing discrepancy between the
experimental values of Ashby {\it et al.}~\cite{AshClavan03} and the
present results.

\begin{acknowledgments}
This work was supported by NNSF of China under Grant No. 10674154.
ZCY was supported by NSERC of Canada and by the computing facilities
of ACEnet, SHARCnet, and WestGrid. JFB was supported by the US NSF
through a grant for the Institute of Theoretical Atomic, Molecular
and Optical Physics at Harvard University and Smithsonian
Astrophysical Observatory. ZCY would also like to thank the Wuhan Institute of
Physics and Mathematics for its hospitality during his visit.
We also thank J. Y. Zhang for helpful discussions.
\end{acknowledgments}



\begin{longtable}{ll}
\caption{\label{g5} Convergence of $\alpha_1$ for the ground state
$2\,^2\!S$ of $^{\infty}$Li, in atomic units.}\\ \hline\hline
\multicolumn{1}{l}{$(N_0,N_P)$}& \multicolumn{1}{c}{$\alpha_1$}\\
\hline
(120, 55)             & 163.92934\\
(256, 138)            & 164.03473\\
(502, 306)            & 164.06958\\
(918, 622)            & 164.09123\\
(1589,1174)           & 164.10058\\
(2625,2091)           & 164.10695\\
(4172,3543)           & 164.11021\\
(6412,5761)           & 164.11154\\
Extrapolated               & 164.112(1)\\
\hline\hline
\end{longtable}


\begin{longtable}{lllc}
\caption{\label{g7} Convergence for the hyperpolarizability of
$^\infty$Li in the ground state $2\,^2\!S$, in atomic units.}\\
\hline\hline \multicolumn{1}{l}{$(N_0,N_S,N_P,N_D)$}&
\multicolumn{1}{l}{$\mathcal{T}(1,0,1)$}&
\multicolumn{1}{c}{$\mathcal{T}(1,2,1)$}&
\multicolumn{1}{c}{$\gamma_0$}\\
\hline
(120,120,55,55)            &--31221.470&  78063.181&   177.926\\
(256,256,138,138)        &--31195.797&  78088.509&   1853.126\\
(502,502,306,306)        &--31177.012&  78100.613&   2958.636\\
(918,918,622,622)        &--31176.318&  78103.456&   3044.300\\
(1589,1589,1174,1174) &--31177.339&  78104.691&   3019.625\\
(2625,2625,2091,2091) &--31176.724&  78102.359&   3004.784\\
(4172,4172,3543,3543) &--31175.663&  78100.447&   3018.655\\
(6412,6412,5761,5761) &--31174.753&  78099.257&   3038.921\\
Extrapolated                  &           &           &   3060(40)\\
\hline\hline
\end{longtable}

\begin{longtable}{clllllll}
\caption{\label{g2} Values of the polarizabilities $\alpha_1$,
$\alpha_2$, and $\alpha_3$, the hyperpolarizability $\gamma_0$, and
the dispersion coefficients $C_6$, $C_8$, and $C_{10}$ for
$^\infty$Li, $^7$Li, and $^6$Li in their ground states $2\,^2\!S$,
in atomic units.}\\
\hline\hline \multicolumn{1}{l}{System}&
\multicolumn{1}{c}{$\alpha_1$}& \multicolumn{1}{c}{$\alpha_2$}&
\multicolumn{1}{c}{$\alpha_3$}& \multicolumn{1}{c}{$\gamma_0$}&
\multicolumn{1}{c}{$C_6$}& \multicolumn{1}{c}{$C_8$}&
\multicolumn{1}{c}{$C_{10}$}\\
\hline
$^\infty$Li & 164.112(1)&1423.263(3)&39649.29(2)&3060(40)&1393.42(5)&83429(1)&73725(2)$\times 10^2$\\
$^7$Li      & 164.161(1)&1423.415(5)&39653.72(3)&2820(40)&1394.05(5)&83456(5)&73742(2)$\times 10^2$\\
$^6$Li      & 164.169(1)&1423.439(4)&39654.46(3)&2780(40)&1394.16(5)&83460(5)&73745(2)$\times 10^2$\\
\hline\hline
\end{longtable}

\begin{longtable}{lllllll}
\caption{\label{g6} Comparison of the polarizabilities $\alpha_1$,
$\alpha_2$, and $\alpha_3$, and the dispersion coefficients $C_6$,
$C_8$, and $C_{10}$ for the ground state $2\,^2\!S$ of $^\infty$Li,
in atomic units. The results from Refs.\cite{DerBabDal01},
\cite{PorDer03}, \cite{Sah07}, \cite{WanSahTim08}, and \cite{JohSafDer08}
were calculated using relativistic methods.}\\
\hline\hline \multicolumn{1}{l}{Reference}&
\multicolumn{1}{c}{$\alpha_1$}& \multicolumn{1}{c}{$\alpha_2$}&
\multicolumn{1}{c}{$\alpha_3$}& \multicolumn{1}{c}{$C_6$}&
\multicolumn{1}{c}{$C_8$}&
\multicolumn{1}{c}{$C_{10}$}\\
\hline
Yan {\it et al.}~\cite{yan}      (1996)    &164.111(2)&1423.266(5)  &39650.49(8)    &1393.39(16) &8.34258(42)$\times 10^4$ &7.3721(1)$\times 10^6$\\
Komasa~\cite{Kom01}             (2001)    &164.11171 &1423.282     &               &            &                         &\\
Derevianko {\it et al.}~\cite{DerBabDal01}
                                 (2001)    &164.0(1)  &             &               &1389(2)     &                         &\\
Porsev and Derevianko~\cite{PorDer03}
                                 (2003)    &          &1424(4)      &3.957$\times 10^4$&         &8.34(4)$\times 10^4$     &7.35$\times 10^6$ \\
Chen and Wang~\cite{CheWan04}   (2004)    &          &1423.48(17)  &39650.96(94)   &            &                         &\\
Cohen and Themelis~\cite{CohThe05}  (2005)    &164.14    &1423.3       &               &            &                         &\\
Zhang {\it et al.}~\cite{ZhaMitBro07}  (2007)    &164.21    &1424.4       &39680          &1394.6      &8.3515$\times 10^4$     &7.3811$\times 10^6$\\
Sahoo~\cite{Sah07}               (2007)    &162.48(56)&1421.37(3.51)&               &            &                         &\\
Wansbeek {\it et al.}~\cite{WanSahTim08}
                                 (2008)    &162.87    &1420         &               &1396(6)     &8.360$\times 10^4$       &\\
Johnson {\it et al.}~\cite{JohSafDer08}
                                 (2008)    &164.084   &1422.73      &39624.2        &            &                         &\\
This work                                  &164.112(1)&1423.263(3)  &39649.29(2)    &1393.42(5)  &8.3429(1)$\times 10^4$  &7.3725(2)$\times 10^6$\\
Molof {\it et al.}~\cite{MolSchMil74} (experiment) (1974)
                                           &164.0(3.4)&             &               &            &                         &\\
Miffre {\it et al.}~\cite{MifJacMuc06} (experiment) (2006)
                                           &164.2(1.1)&             &               &            &                         &\\
\hline\hline
\end{longtable}

\begin{longtable}{ll}
\caption{\label{hyper-results} Values of the hyperpolarizability $\gamma_0$
for
$^\infty$Li in the ground state $2\,^2\!S$,
in atomic units.}\\
\hline\hline \multicolumn{1}{l}{Reference}&\multicolumn{1}{l}{$\gamma_0$}\\
\hline
Pipin and Bishop~\cite{PipBis92}           (1992)    & 3000    \\
Kassimi and Thakkar~\cite{KasTha94,ThaLup06}
                                               (1994)  & 2900(90)\\
Laughlin~\cite{Lau95}                       (1995)  & 3930\\
Jaszu\`{n}ski and Rizzo~\cite{JasRiz96} (1996) &3450\\
Cohen and Themelis~\cite{CohThe05}                (2005)    & 3390\\
Present & 3060(40)\\
\hline\hline
\end{longtable}


\begin{longtable}{llllll}
\caption{\label{g8} Convergence of $\alpha_1$ and $\alpha_1^{(T)}$
for $^{\infty}$Li in $2\,^2\!P$ state, in atomic units.}\\
\hline\hline \multicolumn{1}{l}{$(N_0,N_S,N_{(pp')P},N_D)$}&
\multicolumn{1}{c}{$\alpha_1(S)$}&
\multicolumn{1}{c}{$\alpha_1((pp')P)$}&
\multicolumn{1}{c}{$\alpha_1(D)$}& \multicolumn{1}{c}{$\alpha_1$}&
\multicolumn{1}{c}{$\alpha_1^{(T)}$}\\
\hline
(55,120,36,55)       &--17.181258 &0.063053 &142.611225 &125.493021 &2.951662\\
(138,256,108,138)    &--16.258204 &0.067825 &142.675018 &126.484639 &2.024615\\
(306,502,264,306)    &--15.950171 &0.069135 &142.718553 &126.837517 &1.712883\\
(622,918,568,622)    &--15.871922 &0.069501 &142.729209 &126.926788 &1.633752\\
(1174,1589,1106,1174)&--15.866615 &0.069591 &142.734228 &126.937203 &1.627988\\
(2091,2625,2002,2091)&--15.861376 &0.069641 &142.735430 &126.943695 &1.622654\\
(3543,4172,3413,3543)&--15.860744 &0.069646 &142.736141 &126.945043 &1.621953\\
(5761,6412,3413,5761)&--15.860549 &0.069648 &142.736441 &126.945540 &1.621729\\
Extrapolated                 &            &         &           &126.9458(3)&1.6214(3)\\
\hline\hline
\end{longtable}


\begin{longtable}{cllllll}
\caption{\label{g3} Values of the polarizabilities $\alpha_1$,
$\alpha_1^{(T)}$, $\alpha_2$, and $\alpha_3$, and the
hyperpolarizabilities $\gamma_{0}$ and $\gamma_{2}$
for $^\infty$Li, $^7$Li, and
$^6$Li in $2\,^2\!P$ state, in atomic units.}\\
\hline\hline \multicolumn{1}{l}{System}&
\multicolumn{1}{c}{$\alpha_1$}&
\multicolumn{1}{c}{$\alpha_1^{(T)}$}&
\multicolumn{1}{c}{$\alpha_2$}& \multicolumn{1}{c}{$\alpha_3$}&
\multicolumn{1}{c}{$\gamma_0$}& \multicolumn{1}{c}{$\gamma_2$}\\
\hline
$^\infty$Li & 126.9458(3)&1.6214(3)&4103.165(5)&321138(4)&1.00170(9)$\times 10^7$&--6.2234(8)$\times 10^6$\\
$^7$Li      & 126.9472(5)&1.6351(2)&4102.893(4)&321102(5)&1.00201(9)$\times 10^7$&--6.2252(8)$\times 10^6$\\
$^6$Li      & 126.9474(5)&1.6373(3)&4102.848(4)&321096(5)&1.00206(9)$\times 10^7$&--6.2255(8)$\times 10^6$\\
\hline\hline
\end{longtable}
\begin{longtable}{lllll}
\caption{\label{g9} Comparison of the scalar and tensor dipole
polarizabilities and hyperpolarizabilities for
$^\infty$Li in $2\,^2\!P$ state, in atomic units. The results from \cite{WanSahTim08} and \cite{JohSafDer08} are relativistic.}\\
\hline\hline \multicolumn{1}{l}{Reference}&
\multicolumn{1}{c}{$\alpha_1$}&
\multicolumn{1}{c}{$\alpha_1^{(T)}$}&
\multicolumn{1}{c}{$\gamma_0$}&
\multicolumn{1}{c}{$\gamma_2$}\\
\hline
Themelis and Nicolaides~\cite{TheNic92} (1992)
&135.7      &0.13     &1.10$\times 10^7$      &--6.970$\times 10^6$\\
Pipin and Bishop~\cite{PipBis93} (1993)
&126.844    &1.605    &                       &\\
Cohen and Themelis~\cite{CohThe05} (2005)
&126.4      &1.73     &1.002$\times 10^7$     &--6.21$\times 10^6$\\
Zhang {\it et al.}~\cite{ZhaMitBro07}  (2007)
&126.95     &1.6627   &                       &\\
Wansbeek {\it et al.}~\cite{WanSahTim08} (2008)
&125.20     &5.95     &                       &\\
Johnson {\it et al.}~\cite{JohSafDer08} (2008)
&126.990    &1.59     &                       &\\
This work
&126.9458(3)&1.6214(3)&1.00170(9)$\times 10^7$&--6.2234(8)$\times 10^6$\\
Hunter {\it et al.}~\cite{HunKraBer91} (1991) (experiment)
&126.8(3.4) &         &                       &\\
Windholz {\it et al.}~\cite{WinMusZer92} (1992) (experiment)
&126.87(36) &1.64(4)  &                       &\\
\hline\hline
\end{longtable}

\begin{longtable}{ccccccc}
\caption{\label{g3a} Values of $C_3$, $C_6$, and
$C_8$ for $2\,^2S$--$2\,^2\!P$ of $^\infty$Li,
$^7$Li, and $^6$Li, in atomic units.}\\
\hline\hline
\multicolumn{1}{l}{System}&
\multicolumn{1}{l}{$C_3(M_2=0)$}&
\multicolumn{1}{l}{$C_3(M_2=\pm 1)$}&
\multicolumn{1}{l}{$C_6(M_2=0)$}&
\multicolumn{1}{l}{$C_6(M_2=\pm 1)$}&
\multicolumn{1}{l}{$C_8(M_2=0)$}&
\multicolumn{1}{l}{$C_8(M_2=\pm 1)$}\\
\hline
\multicolumn{7}{c}{$\beta=-1$}\\
$^\infty$Li  &--11.000221(2) &5.500111(1) &2075.40(3) &1406.68(3)  &$990895(5)$ &48564.8(5)\\
$^7$Li       &--11.001853(2) &5.500926(1) &2076.08(7) &1407.15(5)  &$991075(6)$ &48566.4(2)\\
$^6$Li       &--11.002125(2) &5.501062(1) &2076.19(7) &1407.20(2)  &$991104(5)$ &48566.9(4)\\
\multicolumn{7}{c}{$\beta=+1$}\\
$^\infty$Li  &11.000221(2) &--5.500111(1) &2075.40(3) &1406.68(3)  &$274079(2)$ &103044(2)\\
$^7$Li       &11.001853(2) &--5.500926(1) &2076.08(7) &1407.15(5)  &$274128(5)$ &103052(1)\\
$^6$Li       &11.002125(2) &--5.501062(1) &2076.19(7) &1407.20(2)  &$274137(6)$ &103053(1)\\
\hline\hline
\end{longtable}
\begin{longtable}{lrllllll}
\caption{\label{g11} Comparison of $C_3$, $C_6$, and
$C_8$ for the system $^\infty$Li($2\,^2S$)--$^\infty$Li($2\,^2\!P$), in atomic units.}\\
\hline\hline
\multicolumn{1}{l}{Reference}&
\multicolumn{1}{r}{$\beta$}&
\multicolumn{1}{l}{$C_3(M_2=0)$}&
\multicolumn{1}{l}{$C_3(M_2=\pm 1)$}&
\multicolumn{1}{l}{$C_6(M_2=0)$}&
\multicolumn{1}{l}{$C_6(M_2=\pm 1)$}&
\multicolumn{1}{l}{$C_8(M_2=0)$}&
\multicolumn{1}{l}{$C_8(M_2=\pm 1)$}\\
\hline
Marinescu {\it et al.}~\cite{MarDal95} (1995)&--1  &--11.01 &5.503 &2066 &1401 &$9.880\times 10^5$ &$4.756\times 10^4$\\
 &+1 &&&&&$2.705\times 10^5$&$1.021\times 10^5$ \\
Yan {\it et al.}~\cite{yan} (1996) &--1 &--11.000226(15)&5.5001133(74)&2075.05(5)&1406.08(5)&&\\
Zhang {\it et al.}~\cite{ZhaMitBro07} (2007) &--1 &--11.008&5.5041&2076.3&1407.4&$9.9202\times 10^5$&$4.8629\times 10^4$\\
&+1  &&&&&$2.7431\times 10^5$&$1.0316\times 10^5$\\
This work &--1 &--11.000221(2) &5.500111(1) &2075.40(3) &1406.68(3) &$9.90895(5)\times 10^5$ &$4.85648(5)\times 10^4$\\
&+1&&&&&$2.74079(2)\times 10^5$ &$1.03044(2)\times 10^5$\\
\hline\hline
\end{longtable}

\begin{longtable}{llllllllll}
\caption{\label{g10a} Convergence for $\alpha_1$ and $\alpha_1^{(T)}$
of $^{\infty}$Li in $3\,^2\!D$ state, in atomic units.}\\
\hline\hline \multicolumn{1}{l}{$(N_0,N_P,N_{(pd)D},N_F)$}&
\multicolumn{1}{c}{$\alpha_1(P)$}&&
\multicolumn{1}{c}{$\alpha_1((pd)D)$}&&
\multicolumn{1}{c}{$\alpha_1(F)$}&&
\multicolumn{1}{c}{$\alpha_1$}&&
 \multicolumn{1}{c}{$\alpha_1^{(T)}$}
\\
\hline
(138,138,126,132)    &--18857.791600 &&0.059150 &&1916.643488 &&--16941.088962 &&18310.238324\\
(306,306,322,302)    &--16933.479208 &&0.062827 &&1916.940725 &&--15016.475656 &&16385.844685\\
(622,622,714,636)    &--16850.080546 &&0.063886 &&1917.021571 &&--14932.995089 &&16302.423983\\
(1174,1174,1428,1248)&--16845.626886 &&0.064157 &&1917.026639 &&--14928.536090 &&16297.969146\\
(2091,2091,2640,2307)&--16845.378158 &&0.064226 &&1917.035892 &&--14928.278040 &&16297.717843\\
(3543,3543,4587,4051)&--16845.342790 &&0.064242 &&1917.040003 &&--14928.238545 &&16297.681317\\
(5761,5761,4587,6806)&--16845.342870 &&0.064243 &&1917.043303 &&--14928.235324 &&16297.680455\\
\hline\hline
\end{longtable}

\begin{longtable}{lllllllll}
\caption{\label{g10b} Convergence for $\gamma_0$, $\gamma_2$, and
$\gamma_4$ of $^\infty$Li in $3\,^2\!D$ state, in atomic units.}\\
\hline\hline
\multicolumn{1}{l}{$(N_0,N_S,N_P,N_{(pp')P},N_D,N_{(pd)D},N_F,N_{(pf)F},N_G)$}&
\multicolumn{1}{c}{$10^{-12}\,\gamma_0$}&& \multicolumn{1}{c}{$10^{-12}\,\gamma_2$}&&
\multicolumn{1}{c}{$10^{-10}\,\gamma_4$}\\
\hline
(138, 256, 138, 108, 138, 126, 132, 126, 139)    &2.335078929997 &&--2.435345979367 &&10.1097221264\\
(306, 502, 306, 264, 306, 322, 302, 322, 330)    &1.673627590440 &&--1.744707390466 &&7.1916589358\\
(622, 918, 622, 568, 622, 714, 636, 714, 720)    &1.648569007586 &&--1.718605330888 &&7.0875021256\\
(1174,1589,1174,1106,1174,1428,1248,1428,1458)   &1.647221992572 &&--1.717197712093 &&7.0814607892\\
(2091,2625,2091,2002,2091,2640,2307,2640,2769)   &1.647149117913 &&--1.717122212016 &&7.0812009263\\
(3543,4172,3543,3413,3543,4587,4051,4587,4975)   &1.647145129442 &&--1.717120442563 &&7.0814240701\\
\hline\hline
\end{longtable}
\begin{longtable}{clllllll}
\caption{\label{g4} Values of the polarizabilities $\alpha_1$,
$\alpha_1^{(T)}$, $\alpha_2$, and $\alpha_3$, and the
hyperpolarizabilities $\gamma_{0}$, $\gamma_{2}$, and $\gamma_{4}$
for $^\infty$Li, $^7$Li, and
$^6$Li in $3\,^2\!D$ state, in atomic units.}\\
\hline\hline \multicolumn{1}{l}{System}&
\multicolumn{1}{c}{$\alpha_1$}&
\multicolumn{1}{c}{$\alpha_1^{(T)}$}&
\multicolumn{1}{c}{$\alpha_2$}& \multicolumn{1}{c}{$\alpha_3$}&
\multicolumn{1}{c}{$\gamma_0$}& \multicolumn{1}{c}{$\gamma_2$}&
\multicolumn{1}{c}{$\gamma_4$}
\\ \hline
$^\infty$Li &--14928.230(5) &16297.675(5) &158060(10)
&$-1.340902(3)\times 10^8$
&$1.647140(5)\times 10^{12}$ &$-1.717115(5)\times 10^{12}$ &$7.0814(2)\times 10^{10}$\\
$^7$Li      &--14921.330(4) &16291.094(5) &158070(10)
&$-1.339746(5)\times 10^8$
&$1.644875(5)\times 10^{12}$ &$-1.714740(5)\times 10^{12}$ &$7.0700(3)\times 10^{10}$\\
$^6$Li      &--14920.180(6) &16290.000(5) &158070(10)
&$-1.339554(5)\times 10^8$
&$1.644500(4)\times 10^{12}$ &$-1.714345(5)\times 10^{12}$ &$7.0680(5)\times 10^{10}$\\
\hline\hline
\end{longtable}

\begin{longtable}{lllllllllllll}
\caption{\label{g10} Comparison of the scalar and tensor
polarizabilities and hyperpolarizabilities for $^\infty$Li in
$3\,^2\!D$ state, in atomic units. The results from
\cite{WanSahTim08} are relativistic.}
\\
\hline\hline \multicolumn{1}{l}{Reference}&
\multicolumn{1}{l}{$10^{-4}\,\alpha_1$}&
\multicolumn{1}{l}{$10^{-4}\,\alpha_1^{(T)}$}&
\multicolumn{1}{l}{$10^{-5}\,\alpha_2$}&
\multicolumn{1}{l}{$10^{-8}\,\alpha_3$}&
\multicolumn{1}{l}{$10^{-12}\,\gamma_0$}&
\multicolumn{1}{l}{$10^{-12}\,\gamma_2$}&
\multicolumn{1}{l}{$10^{-10}\,\gamma_4$}\\
\hline
Themelis and Nicolaides~\cite{TheNic95} (1995)   &--2.0468       &2.1944        &           &  &4.56&--3.97&19.5\\
Magnier and Aubert-Fr\'{e}con~\cite{MagAubFre02} (2002)     &--1.3950        &1.5324       &           &  & & &\\
Ashby and Wijngaarden~\cite{AshvanWij03} (2003)  &--1.507         &1.642        &           &  & & &\\
Zhang {\it et al.}~\cite{ZhaMitBro07} (2007)             &--1.5044       &1.6414       &1.5786    &--1.3548 & & &\\
Wansbeek {\it et al.}~\cite{WanSahTim08} (2008)       &--1.986        &2.090        &          &         & & &\\
This work                                          &--1.4928230(5)
&1.6297675(5) &1.58060(10)
&--1.340902(3) &1.647140(5) &--1.717115(5) &7.0814(2)\\
Ashby {\it et al.}~\cite{AshClavan03} (2003) (experiment)&--1.513(4)  &1.643(6)  &           & & & &\\
\hline\hline
\end{longtable}

\begin{longtable}{clll}
\caption{\label{g0} Values of oscillator strengths of
$^\infty$Li, $^7$Li, and $^6$Li.}\\
\hline\hline
\multicolumn{1}{l}{System}&
\multicolumn{1}{c}{$2\,^2\!S-2\,^2\!P$}&
\multicolumn{1}{c}{$2\,^2\!P-3\,^2\!D$}&
\multicolumn{1}{c}{$3\,^2\!D-4\,^2\!F$}\\
\hline
$^\infty$Li & 0.7469563(5)&0.6385685(5)&1.0153771(5)\\
$^7$Li      & 0.7469614(4)&0.6385835(5)&1.0154562(5)\\
$^6$Li      & 0.7469623(4)&0.6385858(4)&1.0154695(5)\\
\hline\hline
\end{longtable}

\begin{longtable}{llll}
\caption{\label{g1} Comparison of oscillator strengths of $^\infty$Li.}
\\
\hline\hline
\multicolumn{1}{l}{Reference}
  &\multicolumn{1}{c}{$2\,^2\!S-2\,^2\!P$} &\multicolumn{1}{c}{$2\,^2\!P-3\,^2\!D$} &
   \multicolumn{1}{c}{$3\,^2\!D-4\,^2\!F$} \\
   \hline
Yan and Drake~\cite{YanDra95b}  (1995)   &0.7469572(10)      &0.6385705(30) &  \\
Yan {\it et al.}~\cite{YanTamDra98}        (1998)   &0.7469569396(98) & &  \\
Godefroid {\it et al.}~\cite{GodFroFis01}  (2001)   &0.74690               &0.63853 &  \\
Zhang {\it et al.}~\cite{ZhaMitBro07}          (2007)   &0.7475                  &0.6388         &1.0153\\
Johnson {\it et al.}~\cite{JohSafDer08}      (2008)   &0.746944             &0.638615      &1.015637 \\
This work                                                    &0.7469563(5)       &0.6385685(5) &1.0153771(5) \\
\hline\hline
\end{longtable}
\begin{longtable}{clll}
\caption{\label{sum-rule} Values of $\beta_1$ of
$^\infty$Li, $^7$Li, and $^6$Li in their $2\,^2\!S$, $2\,^2\!P$ and $3\,^2\!D$ states, in atomic units.}\\
\hline\hline \multicolumn{1}{l}{System}&
\multicolumn{1}{c}{$\beta_1(2\,^2\!S)$}&
\multicolumn{1}{c}{$\beta_1(2\,^2\!P)$}&
\multicolumn{1}{c}{$\beta_1(3\,^2\!D)$}\\
\hline
$^\infty$Li & 1196.9696(2)&1614.68(2)&5202428(1)\\
$^7$Li      & 1197.4886(2)&1614.99(2)&5197637(1)\\
$^6$Li      & 1197.5750(2)&1615.04(2)&5196841(1)\\
\hline\hline
\end{longtable}
\appendix
\section{Stark Effect}\label {stark_a}
The Hamiltonian for an atom
in a uniform electric field $\boldsymbol{\cal{E}}=\cal{E}\, \bf{\hat{z}}$ is given by
\begin{eqnarray}
H&=&H_{0}+H'=H_0-\boldsymbol{\cal{E}}\cdot {\bf P}\,, \label
{eq:a1}
\end{eqnarray}
where $H_0$ is the unperturbed Hamiltonian and ${\bf P}$ is the electric dipole moment of the atom:
\begin{eqnarray}
{\bf P}=\sum_{i} q_i {\bf R}_i\,.
\label {eq:a2}
\end{eqnarray}
In the above, $q_i$ is the charge of $i$th particle, ${\bf R}_i$ is its position vector relative
to a laboratory frame, and the summation is over all charged particles inside the atom,
including the nucleus.
Under the perturbation $H'$, the energy eigenvalue and eigenfunction of $H$
can be written in the form
\begin{eqnarray}
E &=& E_0+\Delta E_1 + \Delta E_2 + \Delta E_3 +\Delta E_4+\cdots\,,\\
\Psi &=& \Psi_0 +\Psi_1+\Psi_2\cdots\,,
\label {eq:a3}
\end{eqnarray}
where
\begin{eqnarray}
H_0\Psi_0 &=& E_0\Psi_0\,
\label {eq:a4}
\end{eqnarray}
is the zero-order equation, and $\Delta E_i$ and $\Psi_i$ are the corresponding $i$th-order
corrections. According to the perturbation theory, the energy
corrections can be expressed as
\begin{eqnarray}
\Delta E_1 &=& \langle\Psi_0 |H'|\Psi_0\rangle\,,\\
\Delta E_2 &=& \langle\Psi_0 |H'|\Psi_1\rangle\,,\\
\label {eq:a5_1}
\Delta E_3 &=& \langle\Psi_1|H'|\Psi_1\rangle-\Delta E_1\langle\Psi_1|\Psi_1\rangle\,,\\
\Delta E_4 &=& \langle\Psi_1 |H'|\Psi_2\rangle-\Delta E_2\langle\Psi_1|\Psi_1\rangle
-\Delta E_1 \langle\Psi_1 |\Psi_2\rangle\,.
\label {eq:a5}
\end{eqnarray}
If the state of interest $\Psi_0$ has a fixed parity, as in the case of this work, $\Delta E_1=0$
due to the parity selection rule. $\Psi_1$ and $\Psi_2$ can be expanded in terms of their spectral
representations:
\begin{eqnarray}
\label {eq:a6_1}
|\Psi_1\rangle &=& \sum_n \frac{\langle n|H'|0\rangle}{E_0-E_n}|n\rangle\,,\\
|\Psi_2\rangle &=& \sum_{nk} \frac{\langle n |H'|k\rangle  \langle k |H'|0\rangle}
{(E_0-E_n)(E_0-E_k)}|n\rangle\,,
\label {eq:a6}
\end{eqnarray}
where $|0\rangle\equiv|\Psi_0\rangle$ and $\{E_n,|n\rangle\}$ is a complete set of $H_0$, including the continuum.
Inserting Eq.~(\ref{eq:a6_1}) into Eq.~(\ref{eq:a5_1}) yields $\Delta E_3=0$, also due to the parity consideration.
Thus,
\begin{eqnarray}
\label {eq:a7_1}
\Delta E_2 &=& \sum_n \frac{\langle 0|H'|n\rangle \langle n|H'|0\rangle}{E_0-E_n}\,,\\
\Delta E_4 &=& \sum_{kmn}\bigg[\frac{1}{(E_0-E_m)(E_0-E_n)(E_0-E_k)}
-\delta(n,0)\frac{1}{(E_0-E_m)(E_0-E_k)^2}\bigg]\nonumber\\
&&\times \langle 0|H'|m\rangle \langle m|H'|n\rangle \langle n|H'|k\rangle\langle k|H'|0\rangle\,.
\label {eq:a7}
\end{eqnarray}

\subsection{$\Delta E_2$}\label {deltae2}
Let us first consider the operator $H'|n\rangle\langle n|H'$ in Eq.~(\ref{eq:a7_1}). Using spherical
tensor operator technique, we have the following decomposition
\begin{eqnarray}
H'|n\rangle\langle n|H' &=&\sum_{Kq} (-1)^{K+q}[P^{(1)}\otimes\lambda_nP^{(1)}]_q^{(K)}
[\mathcal {E}^{(1)}\otimes \mathcal {E}^{(1)} ]_{-q}^{(K)}\,,
\label {eq:a8}
\end{eqnarray}
where $\lambda\equiv |n\rangle\langle n|$. Since $\boldsymbol{\cal{E}}$ is along the $z$-axis,
only the $q=0$ component survives in the above equation. Thus,
\begin{eqnarray}
H'|n\rangle\langle n|H' &=&\sum_K(-1)^K\sqrt{2K+1}\sum_{q_1q_2}
\left (
\begin{matrix}
  1 & 1 & K \\
  q_1 & q_2 & 0 \\
\end{matrix}
\right)
P_{q_1}^{(1)}\lambda_nP_{q_2}^{(1)}
[\mathcal {E}^{(1)}\otimes \mathcal {E}^{(1)} ]_{0}^{(K)}\,.
\label {eq:a9}
\end{eqnarray}
After substituting the Eq~(\ref{eq:a9}) into Eq~(\ref{eq:a7_1}), one has
\begin{eqnarray}
\Delta E_2 &=& \sum_{nK} (-1)^K\sqrt{2K+1}\sum_{q_1q_2}
\left (
\begin{matrix}
  1 & 1 & K \\
  q_1 & q_2 & 0 \\
\end{matrix}
\right)
\frac{\langle 0 |P_{q_1}^{(1)}|n\rangle \langle n |P_{q_2}^{(1)}|0\rangle}
{E_0-E_n}[\mathcal {E}^{(1)}\otimes \mathcal {E}^{(1)} ]_{0}^{(K)}\,.
\label {eq:a10}
\end{eqnarray}
To be specific, let us write out explicitly the angular momentum quantum numbers in the initial and intermediate
states:
\begin{eqnarray}
|0\rangle &=& |n_0 LM\rangle \,,\\
|n\rangle &=& |nL_a M_a\rangle \,,
\label {eq:a11}
\end{eqnarray}
where $n_0$ and $n$ are the corresponding principal quantum numbers. Then, the summation over $n$ in
Eq~(\ref{eq:a10}) actually means the summation over $\{n,L_a,M_a\}$. By applying the Wigner-Eckart theorem
for the irreducible tensor operator $P_\mu^{(1)}$, one can recast Eq~(\ref{eq:a10}) into the following form
\begin{eqnarray}
\Delta E_2 &=& \sum_{nL_aK} (-1)^K\sqrt{2K+1}\,
\frac{\langle n_0L \|P^{(1)}\|nL_a\rangle \langle nL_a \|P^{(1)}\|n_0L\rangle}
{E_{n_0}(L)-E_n(L_a)}[\mathcal {E}^{(1)}\otimes \mathcal {E}^{(1)} ]_{0}^{(K)}\,A\,,
\label {eq:a12}
\end{eqnarray}
where $E_{n_0}(L)$ and $E_n(L_a)$ stand for $E_0$ and $E_n$ respectively in Eq.~(\ref{eq:a10}), and
\begin{eqnarray}
A &=&\sum_{M_aq_1q_2}(-1)^{L-M+L_a-M_a}
\left (
\begin{matrix}
L&1&L_a\\
-M&q_1&M_a\\
\end{matrix}
\right )
\left(
\begin{matrix}
L_a&1&L\\
-M_a&q_2&M\\
\end{matrix}
\right )
\left (
\begin{matrix}
1&1&K\\
q_1&q_2&0\\
\end{matrix}
\right )\,.
\label {eq:a13}
\end{eqnarray}
The quantity
$A$ can be simplified by using the standard graphical method of dealing with angular momentum~\cite{Zar88},
\begin{eqnarray}
A &=& (-1)^{L-M}
\left(
\begin{matrix}
L&L&K\\
-M&M&0
\end{matrix}
\right)
\left\{
\begin{matrix}
1&1&K\\
L&L&L_a
\end{matrix}
\right\}\,,
\label {eq:a14}
\end{eqnarray}
provided $K$ is an integer, and note that
\begin{eqnarray}
[\mathcal {E}^{(1)}\otimes \mathcal {E}^{(1)} ]_{0}^{(K)}&=&
\sqrt{2K+1}\,\sum_{q_1q_2}
\left(
\begin{matrix}
1&1&K\\
q_1&q_2&0
\end{matrix}
\right)
\mathcal{E}_{q_1}^{(1)}\mathcal{E}_{q_2}^{(1)}
= \sqrt{2K+1}
\left(
\begin{matrix}
1&1&K\\
0&0&0
\end{matrix}
\right)
\mathcal{E}^2\,.
\label {eq:a15}
\end{eqnarray}
Therefore, the second-order correction can be written
\begin{eqnarray}
\Delta E_2 &=& -\mathcal{E}^2\,\sum_{nL_a}
\frac{\langle n_0L \|P^{(1)}\|nL_a\rangle\langle nL_a \|P^{(1)}\|n_0L\rangle}
{E_n(L_a)-E_{n_0}(L)}\nonumber\\
&\times&\sum_K(2K+1)(-1)^{L-M}
\left(
\begin{matrix}
1&1&K\\
0&0&0
\end{matrix}
\right)
\left(
\begin{matrix}
L&L&K\\
-M&M&0
\end{matrix}
\right)
\left\{
\begin{matrix}
1&1&K\\
L&L&L_a
\end{matrix}
\right\}\,.
\label {eq:a16}
\end{eqnarray}
Since
\begin{eqnarray}
\label {eq:a17_1}
(-1)^{L-M}
\left (
\begin{matrix}
L&L&0\\
-M&M&0
\end{matrix}
\right )
&=&(-1)^{2L}\frac{1}{\sqrt{2L+1}}\,,\\
(-1)^{L-M}
\left (
\begin{matrix}
L&L&2\\
-M&M&0
\end{matrix}
\right)
&=&(-1)^{2L}\frac{3M^2-L(L+1)}{\sqrt{(2L+3)(L+1)(2L+1)L(2L-1)}}\,,  \ \ L\ge 1\,,
\label {eq:a17}
\end{eqnarray}
and also~\cite{Zar88}
\begin{eqnarray}
\label {eq:a18_1}
P_\mu^{(1)} &=& \sqrt{\frac{4\pi}{3}}\sum_i q_i R_i Y_{1 \mu}(\hat{\bf R}_i)\,,\\
\langle n L_a \| P^{(1)} \| n_0 L\rangle &=&
(-1)^{L-L_a}\langle n_0L \| P^{(1)} \| nL_a\rangle^*\,,
\label {eq:a18}
\end{eqnarray}
the second-order energy correction can finally be expressed in the form
\begin{eqnarray}
\Delta E_2 &=&
-\frac{{\cal{E}}^2}{2}(\alpha_1+\alpha_1^{(T)}g_2(L,M))\,.
 \label {eq:a19}
\end{eqnarray}
In the above, $g_2(L,M)$ is the only $M$-dependent part:
\begin{eqnarray}
 g_2(L,M) =\left\{ \begin{array}{ll}
         0, & \mbox{if $L=0$, $\frac{1}{2}$} \\
        \dfrac{3M^2-L(L+1)}{L(2L-1)}, & \mbox{otherwise}\end{array} \right.
\label {eq:a20}
\end{eqnarray}
and $\alpha_1$ and $\alpha_1^{(T)}$ are, respectively,  the scalar and tensor dipole
polarizabilities
\begin{eqnarray}
\label{eq:a21_1}
\alpha_1 &=& \sum_{L_a} \alpha_1(L_a)\,, \\
 \alpha_1^{(T)} &=& \sum_{L_a}
W(L,L_a)\alpha_1(L_a)\,,
\label {eq:a21}
\end{eqnarray}
where
\begin{eqnarray}
\alpha_1(L_a) &=& \frac{8\pi}{9(2L+1)}\sum_n \frac{|\langle n_0 L
\|T_1\| nL_a\rangle |^2}{E_n(L_a)-E_{n_0}(L)}\,,
 \label {eq:a22}
\end{eqnarray}
with $T_1=\sum_i q_iR_i Y_{10}({\bf {\hat{R}}}_i)$, and
\begin{eqnarray}
W(L,L_a) &=& (-1)^{L+L_a}\sqrt{\frac{30(2L+1)L(2L-1)}{(2L+3)(L+1)}}
\left\{%
\begin{matrix}
  1 & 1 & 2 \\
  L & L & L_a \\
\end{matrix}%
\right\}\,.
 \label {eq:a23}
\end{eqnarray}

\subsection{$\Delta E_4$}\label {deltae4}

According to Eq.~(\ref{eq:a9}), the fourth-order energy correction of Eq.~(\ref{eq:a7}) can be written as
\begin{eqnarray}
\Delta E_4 &=&\sum_{kmn} t(k,m,n) \sum_{K_1K_2}(-1)^{K_1+K_2}(K_1,K_2)^{1/2}\sum_{q_1q_2q_3q_4}
\left(
\begin{matrix}
1&1&K_1\\
q_1&q_2&0
\end{matrix}
\right)
\left(
\begin{matrix}
1&1&K_2\\
q_3&q_4&0
\end{matrix}
\right)\nonumber\\
&\times&
\langle 0 |P_{q_1}^{(1)} \lambda_m P_{q_2}^{(1)} \lambda_n P_{q_3}^{(1)} \lambda_k P_{q_4}^{(1)}
|0\rangle [\mathcal {E}^{(1)}\otimes \mathcal {E}^{(1)} ]_{0}^{(K_1)}
[\mathcal {E}^{(1)}\otimes \mathcal {E}^{(1)} ]_{0}^{(K_2)}\,,
\label {eq:a28}
\end{eqnarray}
where the notation $(a,b)=(2a+1)(2b+1)$,
\begin{eqnarray}
t(k,m,n) &=& \frac{1}{(E_0-E_m)(E_0-E_n)(E_0-E_k)}-\delta(n,0)\frac{1}{(E_0-E_m)(E_0-E_k)^2}\,,
\label {eq:a29}
\end{eqnarray}
and $\lambda_m=|m\rangle\langle m|$, {\it etc.}.
Writing out the all angular momentum quantum numbers explicitly
\begin{eqnarray}
|0\rangle &=& |n_0 LM\rangle\,,\\
|m\rangle &=& |m L_aM_a\rangle\,,\\
|n\rangle &=& |n L_bM_b\rangle\,,\\
|k\rangle &=& |k L_cM_c\rangle\,,
\label {eq:a30}
\end{eqnarray}
and applying the Wigner-Eckart theorem, we have
\begin{eqnarray}
\Delta E_4 &=& \sum_{kmn}\sum_{L_aL_bL_c}t(k,m,n)\sum_{K_1K_2}(-1)^{K_1+K_2}(K_1,K_2)^{1/2}
[\mathcal {E}^{(1)}\otimes \mathcal {E}^{(1)} ]_{0}^{(K_1)}
[\mathcal {E}^{(1)}\otimes \mathcal {E}^{(1)} ]_{0}^{(K_2)}\nonumber\\
&\times&
\langle n_0L\|P^{(1)}\|mL_a\rangle
\langle mL_a\|P^{(1)}\|nL_b\rangle
\langle nL_b\|P^{(1)}\|kL_c\rangle
\langle kL_c\|P^{(1)}\|n_0L\rangle\, B\,,
\label {eq:a31}
\end{eqnarray}
where $B$ (which contains all the angular coefficients) is
\begin{eqnarray}
B&=& \sum_{q_1q_2q_3q_4}\sum_{M_aM_bM_c}(-1)^{L-M+L_a-M_a+L_b-M_b+L_c-M_c}
\left (
\begin{matrix}
1&1&K_1\\
q_1&q_2&0
\end{matrix}
\right )
\left (
\begin{matrix}
1&1&K_2\\
q_3&q_4&0
\end{matrix}
\right )
\left (
\begin{matrix}
L&1&L_a\\
-M&q_1&M_a
\end{matrix}
\right )\nonumber\\
&\times&
\left (
\begin{matrix}
L_a&1&L_b\\
-M_a&q_2&M_b
\end{matrix}
\right )
\left (
\begin{matrix}
L_b&1&L_c\\
-M_b&q_3&M_c
\end{matrix}
\right )
\left (
\begin{matrix}
L_c&1&L\\
-M_c&q_4&M
\end{matrix}
\right )\,.
\label {eq:a32}
\end{eqnarray}
Using the graphical method, $B$ can be simplified into
\begin{eqnarray}
B &=&(-1)^{L-M}
\left \{
\begin{matrix}
1&1&K_1\\
L&L_b&L_a
\end{matrix}
\right \}
\left \{
\begin{matrix}
1&1&K_2\\
L&L_b&L_c
\end{matrix}
\right \}
\sum_\lambda(2\lambda+1)
\left (
\begin{matrix}
K_1&K_2&\lambda\\
0&0&0
\end{matrix}
\right )
\left (
\begin{matrix}
L&L&\lambda\\
-M&M&0
\end{matrix}
\right )
\left \{
\begin{matrix}
K_2&K_1&\lambda\\
L&L&L_b
\end{matrix}
\right \}\,,
\label {eq:a33}
\end{eqnarray}
provided $K_1$ and $K_2$ are integers.
From Eqs.~(\ref{eq:a15}), (\ref{eq:a18_1}), and (\ref{eq:a18}), one can further
write Eq.~(\ref{eq:a31}) in the form
\begin{eqnarray}
\Delta E_4 &=& -\mathcal{E}^4 \left(\frac{4\pi}{3}\right )^2\sum_{L_aL_bL_c}
\mathcal {T}(L_a,L_b,L_c)
\sum_{\lambda}
(-1)^{L-M}
\left (
\begin{matrix}
L&L&\lambda\\
-M&M&0
\end{matrix}
\right )
\sum_{K_1K_2}(\lambda,K_1,K_2)
\left (
\begin{matrix}
1&1&K_1\\
0&0&0
\end{matrix}
\right )
\left (
\begin{matrix}
1&1&K_2\\
0&0&0
\end{matrix}
\right )\nonumber\\
&\times&
\left (
\begin{matrix}
K_1&K_2&\lambda\\
0&0&0
\end{matrix}
\right )
\left \{
\begin{matrix}
1&1&K_1\\
L&L_b&L_a
\end{matrix}
\right \}
\left \{
\begin{matrix}
1&1&K_2\\
L&L_b&L_c
\end{matrix}
\right \}
\left \{
\begin{matrix}
K_2&K_1&\lambda\\
L&L&L_b
\end{matrix}
\right \}\,,
\label {eq:a34}
\end{eqnarray}
where
\begin{eqnarray}
\mathcal{T}(L_a,L_b,L_c) &=& \sum_{kmn}\frac{\langle
n_0L\|T_1\|mL_a\rangle \langle mL_a\|T_1\|nL_b\rangle \langle
nL_b\|T_1\|kL_c\rangle \langle kL_c\|T_1\|n_0L\rangle }
{[E_k(L_c)-E_{n_0}(L)][E_m(L_a)-E_{n_0}(L)][E_n(L_b)-E_{n_0}(L)]}\nonumber
\\
&-& \delta(L_b,L)(-1)^{2L-L_a-L_c}
\sum_m
\frac{|\langle n_0L\|T_1\|mL_a\rangle|^2}
{[E_m(L_a)-E_{n_0}(L)]} \sum_k
\frac{|\langle n_0L\|T_1\|kL_c\rangle|^2 }{[E_k(L_c)-E_{n_0}(L)]^2} \,.
\label {eq:a35}
\end{eqnarray}
According to the property of $3j$ symbol, the possible values for $\lambda$ are 0, 2, and 4.
Also,
\begin{eqnarray}
&&(-1)^{L-M}
\left (
\begin{matrix}
L&L&4\\
-M&M&0
\end{matrix}
\right )= \nonumber\\
&&\frac
{2(-1)^{2L}[3(5M^2-L^2-2L)(5M^2+1-L^2)-10M^2(4M^2-1)]}
{\sqrt{(2L+5)(2L+4)(2L+3)(2L+2)(2L+1)(2L)(2L-1)(2L-2)(2L-3)}}\,, \ \ L\ge 2\,.
\label {eq:a36}
\end{eqnarray}
Together with Eqs.~(\ref{eq:a17_1}) and (\ref{eq:a17}), the fourth-order correction can finally
be expressed in the form
\begin{eqnarray}
\Delta E_4 &=& -\frac{{\cal{E}}^4}{24}(\gamma_0+\gamma_2\,g_2(L,M))
+\gamma_4\,g_4(L,M))\,,
 \label {eq:a37}
\end{eqnarray}
where $g_2(L,M)$ is defined in Eq.~(\ref{eq:a20}), and $g_4(L,M)$
is given by
\begin{eqnarray}
 g_4(L,M) =\left\{ \begin{array}{ll}
         0, & \mbox{if $L\le \frac{3}{2}$, } \\
        \dfrac{3(5M^2-L^2-2L)(5M^2+1-L^2)-10M^2(4M^2-1)}
        {L(2L-1)(2L-2)(2L-3)}, & \mbox{otherwise}\end{array} \right.
\label {eq:a38}
\end{eqnarray}
In Eq.~(\ref{eq:a37}), $\gamma_0$ is the scalar
hyperpolarizability, and $\gamma_2$ and $\gamma_4$ are the tensor
hyperpolarizabilities, which can be written as
\begin{eqnarray}
\gamma_{0}&=&(-1)^{2L}\frac{128\pi^{2}}{3}\frac{1}{\sqrt{2L+1}}\sum_{L_{a}L_{b}L_{c}}
\mathcal {G}_{0}(L,L_{a},L_{b},L_{c})\mathcal{T}(L_{a},L_{b},L_{c}),
\label {eq:a39}
\end{eqnarray}
\begin{eqnarray}
\gamma_{2}&=&(-1)^{2L}\frac{128\pi^{2}}{3}\sqrt{\frac{L(2L-1)}{(2L+3)(L+1)(2L+1)}}
\sum_{L_{a}L_{b}L_{c}}
\mathcal{G}_{2}(L,L_{a},L_{b},L_{c})\mathcal{T}(L_{a},L_{b},L_{c}),
\label {eq:a40}
\end{eqnarray}
\begin{eqnarray}
\gamma_{4}&=&(-1)^{2L}\frac{128\pi^{2}}{3}
\sqrt{\frac{L(2L-1)(L-1)(2L-3)} {(2L+5)(L+2)(2L+3)(L+1)(2L+1)}}
\sum_{L_{a}L_{b}L_{c}}
\mathcal{G}_{4}(L,L_{a},L_{b},L_{c})\mathcal{T}(L_{a},L_{b},L_{c}),
\label {eq:a41}
\end{eqnarray}
where
\begin{eqnarray}
\mathcal{G}_{\Lambda}(L,L_{a},L_{b},L_{c})&=&\sum_{K_{1}K_{2}}(\Lambda,K_{1},K_{2})
\left(
\begin{matrix}
  1 & 1 & K_{1} \\
  0 & 0 & 0 \\
\end{matrix}
\right)
\left(
\begin{matrix}
  1 & 1 & K_{2} \\
  0 & 0 & 0 \\
\end{matrix}
\right)\left(
\begin{matrix}
  K_{1} & K_{2} & \Lambda \\
  0 & 0 & 0\\
\end{matrix}
\right)
\left\{
\begin{matrix}
  1 & 1 & K_{1} \\
  L & L_{b} & L_{a} \\
\end{matrix}
\right\}\nonumber\\
&\times&
\left\{
\begin{matrix}
  1 & 1 & K_{2} \\
  L & L_{b} & L_{c} \\
\end{matrix}
\right\}
\left\{
\begin{matrix}
  K_{2} & K_{1} & \Lambda \\
  L & L & L_{b} \\
\end{matrix}
\right\}\,. \label {eq:a42}
\end{eqnarray}

\end{document}